\documentclass[ALICE,manyauthors]{cernphprep}

\newcommand{\nineH}{$\sqrt{s}~=~0.9$~Te\kern-.1emV}
\newcommand{\seven}{$\sqrt{s}~=~7$~Te\kern-.1emV}
\newcommand{\twoH}{$\sqrt{s}~=~0.2$~Te\kern-.1emV}

\newcommand{\GeVc}{Ge\kern-.1emV/$c$}
\newcommand{\MeVc}{Me\kern-.1emV/$c$}
\newcommand{\TeV}{Te\kern-.1emV}
\newcommand{\GeV}{Ge\kern-.1emV}
\newcommand{\MeV}{Me\kern-.1emV}
\newcommand{\GeVmass}{Ge\kern-.2emV/$c^2$}
\newcommand{\MeVmass}{Me\kern-.2emV/$c^2$}

\newcommand{\pp}           {pp}
\newcommand{\ppbar}        {\mbox{$\mathrm {p\overline{p}}$}}
\newcommand{\XeXe}         {\mbox{Xe--Xe}}
\newcommand{\PbPb}         {\mbox{Pb--Pb}}
\newcommand{\pA}           {\mbox{pA}}
\newcommand{\pPb}          {\mbox{p--Pb}}
\newcommand{\AuAu}         {\mbox{Au--Au}}

\newcommand{\dndy}         {\ensuremath{\mathrm{d}N_\mathrm{ch}/\mathrm{d}y}}
\newcommand{\dndeta}       {\ensuremath{\mathrm{d}N_\mathrm{ch}/\mathrm{d}\eta}}
\newcommand{\dNdeta}       {\ensuremath{\langle\dndeta\rangle}}

\newcommand{\s}            {\ensuremath{\sqrt{s}}}

\newcommand{\snn}          {\ensuremath{\sqrt{\it{s}_{\mathrm{NN}}}}}
\newcommand{\snnbf}        {\ensuremath{\mathbf{{\sqrt{\it{s}_{\mathbf{NN}}}}}}}

\newcommand{\Npart}        {\ensuremath{N_\mathrm{part}}}
\newcommand{\Npartcore}    {\ensuremath{N_\mathrm{part}^\mathrm{core}}}
\newcommand{\Npartcorona}  {\ensuremath{N_\mathrm{part}^\mathrm{corona}}}
\newcommand{\Ncpart}       {\ensuremath{N_\text{q-part}}}
\newcommand{\Ncn}          {\ensuremath{N_\text{q}}}
\newcommand{\avNpart}      {\ensuremath{\langle N_\mathrm{part} \rangle}}
\newcommand{\avNcpart}     {\ensuremath{\langle N_\text{q-part} \rangle}}

\newcommand{\Ncoll}        {\ensuremath{N_\mathrm{coll}}}

\newcommand{\dNdetape}     {\ensuremath{\frac{2}{\avNpart}\dNdeta}} 

\newcommand{\Ntot}         {\ensuremath{N_\mathrm{ch}^\mathrm{tot}} }

\newcommand{\abs}[1]       {\ensuremath{\left|#1\right|}}

\usepackage[usenames]{color}
\usepackage{multirow}
\usepackage[normalem]{ulem} 
\usepackage{units}

\RequirePackage{color}\definecolor{RED}{rgb}{1,0,0}\definecolor{BLUE}{rgb}{0,0,1}

\usepackage{fancyhdr}
\usepackage[comma,square,numbers,sort&compress]{natbib}
\usepackage{hyperref}
\usepackage{lineno}

\begin{document}%
\newlength{\figlen}
\setlength{\figlen}{0.75\textwidth}

\begin{titlepage}
  \PHyear{2018}
  \PHnumber{120}      
  \PHdate{9 May}  
  %

\title{Centrality and pseudorapidity dependence of the charged-particle
multiplicity density in Xe--Xe collisions at \snnbf\ = \unit[5.44]{TeV}}
\ShortTitle{\dndeta\ in \XeXe\  at \snn\ = \unit[5.44]{TeV}}   

\Collaboration{ALICE Collaboration\thanks{See Appendix~\ref{app:collab} for the list of collaboration members}}
\ShortAuthor{ALICE Collaboration} 

\begin{abstract}
In this Letter, the ALICE Collaboration presents the first measurements of the
charged-particle multiplicity density, \dndeta,
and total charged-particle multiplicity, $N^\mathrm{tot}_\mathrm{ch}$, in \XeXe\ collisions at a
centre-of-mass energy per nucleon--nucleon pair of \snn~=~\unit[5.44]{TeV}.
The measurements are performed as a function of collision centrality over a
wide pseudorapidity range of $-3.5 < \eta < 5$.
The values of \dndeta\ at mid-rapidity and \Ntot\ for central collisions, normalised to
the number of nucleons participating in the collision (\Npart) as a function of \snn\,
follow the trends established in previous heavy-ion measurements.
The same quantities are also found to increase as a function of \Npart,
and up to the 5\% most central collisions the trends are the same
as the ones observed in \PbPb\ at a similar
energy. For more central collisions,
the \XeXe\ scaled multiplicities exceed
those in \PbPb\ for a similar \Npart.  The results are
compared to phenomenological models and theoretical calculations based
on different mechanisms for particle production in nuclear collisions.
All considered models describe the data reasonably well within 15\%.

\end{abstract}
\end{titlepage}
\setcounter{page}{2}

%
%
\section{Introduction}
\label{sec:intro}
A plasma of strongly
interacting quarks and gluons is formed in the hot and dense nuclear matter created in
ultra-relativistic heavy-ion collisions~\cite{Karsch:2001cy,Muller:2012zq}.
The multiplicity of
charged particles produced in the collisions is a key observable to
characterise the properties of the matter created in these collisions,
as the overall particle production is related to the initial energy density.
Nuclei are extended objects and the degree of geometrical overlap
between them in the collision, expressed in terms of the impact parameter ($b$),
varies. Since $b$ is not directly measurable, an experimental proxy of centrality
is used to characterise the amount of nuclear overlap in the collisions.
Typical features related to the collision
centrality are the number of nucleons participating in the collision,
\Npart, and the number of binary nucleon-nucleon collisions, \Ncoll,
among the participant nucleons.
Collisions of nuclei of different sizes lead to different \Npart\ and \Ncoll\
for similar relative nuclear overlap.
The study of the production of charged particles with different collision systems
and at various collision energies can help shed light on the role of the initial energy density
and the production mechanism of final-state particles.

Previous measurements of the system-size dependence of the
charged-particle pseudorapidity density (\dndeta) were performed at
RHIC, comparing Au--Au and Cu--Cu collisions at various centre-of-mass energies~\cite{Alver:2010ck}.
The ALICE, ATLAS and CMS Collaborations at the LHC have previously reported on
\dndeta\ in Pb--Pb collisions at
\snn~=~\unit[2.76]{TeV}~\cite{PhysRevLett.106.032301,Abbas:2013bpa,Aad2012363,Chatrchyan:2011aa}
and \unit[5.02]{TeV}~\cite{Adam:2015ptt,Adam:2016ddh}.
The dependence of the charged-particle density averaged at mid-rapidity
($|\eta|$ $<$ 0.5) \dNdeta\ over the centre-of-mass energy shows
a steeper increase in central heavy-ion collisions than in
proton--proton (\pp) and proton--nucleus (\pA) collisions.
The values of \dNdeta, normalised by the number of nucleon
pairs participating in the collision, increase faster than linearly with \Npart.
No significant differences between the shapes of the \Npart\ dependence
for the different collision energies were observed.

In this Letter, the ALICE Collaboration presents the first measurement
of the production of charged, primary particles in \XeXe\ collisions at
\snn~=~\unit[5.44]{TeV}. The size of the \XeXe\ system is intermediate between
previously studied systems at the LHC, \PbPb\
\cite{PhysRevLett.106.032301,Abbas:2013bpa,Adam:2015ptt,Adam:2016ddh} being the largest
and \pPb~and pp~\cite{LongMultiPaper,PhysRevLett.110.032301} the smallest.
The charged-particle pseudorapidity density is presented
over the interval $-3.5< \eta <5$ and as a function of the collision centrality.
The mid-rapidity values normalised by the number of participating nucleon--nucleon pairs
are also reported.
The results are also compared
with measurements at lower collision energies and with theoretical
calculations.

\section{Experimental setup}
\label{sec:exper}
The data were recorded with the ALICE apparatus in 6 hours of stable
data-taking with $^{129}$Xe beams (16 bunches per beam) colliding at
\snn~=~\unit[5.44]{TeV} in October 2017.  The data were collected with a
reduced magnetic field of \unit[0.2]{T} (as compared to the nominal value of \unit[0.5]{T}) in the
ALICE solenoid magnet. The performance and a detailed description of ALICE can be found elsewhere~\cite{Abelev:2014ffa}.
In the following, the detector elements relevant to this
analysis are briefly described.

The innermost part of the tracking system of ALICE is the Silicon
Pixel Detector (SPD)~\cite{1748-0221-4-03-P03023}
which consists of two cylindrical layers of
hybrid silicon pixel assemblies.
The inner and outer SPD layers are placed at radii of 3.9 and \unit[7.6]{cm}
from the interaction point and cover $|\eta|<2$ and $|\eta|<1.4$, respectively.
The Forward Multiplicity Detector
(FMD)~\cite{Christensen:2007yc, Cortese:781854}
consists of three sets of silicon strip
sensors, covering the pseudorapidities $-3.5<\eta<-1.8$ and
$1.8<\eta<5$.
The FMD records the energy deposited by
charged particles impinging the detector.
The V0 detector~\cite{Abbas:2013taa, Cortese:781854} is used for
triggering and centrality classification.
It consists of two sub-detectors, V0-A and V0-C,
covering the pseudorapidity regions $2.8<\eta<5.1$ and $-3.7<\eta<-1.7$,
respectively.  The V0 has a timing resolution better than \unit[1]{ns},
allowing its fast signals to be combined in a programmable logic to
reject beam-induced background events while ensuring maximum efficiency
for the selection of collision events.
The Zero-Degree Calorimeters (ZDCs)~\cite{PUDDU2007397}
are located at a distance of \unit[112.5]{m} from the interaction point along the beam line,
on either side of the experiment.
They measure the energy of spectator (non--interacting)
nucleons. The ZDCs are also used for triggering and provide timing information
used to select collisions occurring in the interaction point region.

\section{Data sample and analysis method}
\label{sec:data}

The hadronic interaction rate in ALICE
was about \unit[150 (80)]{Hz} at the beginning (end) of the data-taking.
The magnetic field of \unit[0.2]{T}, reduced as compared to normal \PbPb\ settings (\unit[0.5]{T})
increases the acceptance for low-momentum particles,
thus enhancing the acceptance of the V0 system
for electromagnetic (EM) interactions, which constitute a background for this analysis.
In order to suppress this source of contamination, the minimum bias
interaction trigger required a signal in each of the V0 sub-detectors in coincidence with a signal
in each of the two neutron ZDCs. It was verified by means of a set of control triggers
that such a trigger is fully efficient for hadronic interactions in the 0--90\% centrality range.
In addition, beam-background interactions are removed using the V0 and the ZDC timing
information.
The interaction probability per bunch-crossing was sufficiently small
that the chance of two hadronic interactions occurring within the
integration time of the involved detectors,
so-called pileup events, was negligible.  A total of about 1 million
hadronic collisions are used in this analysis.

The classification of collisions into centrality classes uses the sum of the
amplitudes of the signals in the $\mbox{V0-A}$ and $\mbox{V0-C}$
detectors.
A model of particle production, based on a Glauber description~\cite{Alver:2008aq,Loizides:2014vua},
is fitted to the V0 amplitude distribution~\cite{PhysRevC.88.044909}.
The number of particles in the V0 detector is calculated with a
two-component model for the number of sources given by
\begin{equation}
\label{eqtwocomponent}
N_\mathrm{sources} = f\times N_\mathrm{part} + (1-f)\times N_\mathrm{coll} \quad,
\end{equation}
where $f$ constrains the relative contributions of \Npart\ and \Ncoll,
coupled to a particle production model for each source parameterised
by the negative binomial distribution (NBD).
In the Glauber calculation, the nuclear density
for $^{129}$Xe is described by a Woods-Saxon distribution for a
deformed nucleus

\begin{equation}
\rho(r, \vartheta) = \rho_{0} \frac{1}{1 + \exp \left( \frac{r - R(\vartheta)}{a} \right)}\quad.
\end{equation}
The parameter $\rho_{0}$ is the nucleon density, which provides the
overall normalisation. The nuclear skin thickness is $a=0.59\pm0.07\,\mathrm{fm}$ \cite{Tsukada:2017llu}.
The nuclear radius $R$ is parametrised as a
function of the polar angle $\vartheta$ by $R(\vartheta) = R_0
[1+\beta_2Y_{20}(\vartheta)]$, where $R_0$ is the average radius and the
Legendre polynomial $Y_{20}$ describes the nucleus deformation for an
axially symmetric case with no dependence on the azimuthal
angle. For the average radius we used $R_0=5.4\pm0.1\,\mathrm{fm}$, scaling the results for $^{132}$Xe reported in
\cite{Tsukada:2017llu} by the atomic mass number (A) dependence of the
radius, namely $(129/132)^{1/3}$~\cite{Loizides:2014vua}.   The deformation parameter
$\beta_2=0.18 \pm 0.02$ is obtained by linearly interpolating the values
measured for the Xe A-even isotopes from 124 to 136~\cite{ALICE-PUBLIC-2018-003}.
In the Glauber model calculation,
the orientation of the spheroid symmetry axis is randomly sampled.  For
\snn~=~\unit[5.44]{TeV} collisions, an inelastic nucleon--nucleon cross section
of $68.4 \pm 0.5\,\mathrm{mb}$, obtained by logarithmic interpolation
of cross section measurements with respect to collision energies in pp collisions
\cite{Loizides:2017ack}, is used.  The NBD-Glauber fit provides a good
description of the observed V0 amplitude in the region corresponding
to the top 90\% of the hadronic cross section, where the
effects of trigger inefficiency and contamination by EM processes are
negligible.
The average numbers of participants \avNpart\
reported in Tab.~\ref{tab:data} are estimated from the Glauber model
imposing the same cuts applied to the data on the simulated V0 response.
One should note that the centrality selection based on the V0 amplitude induces
a bias on the measured \dNdeta.
This leads to a \dNdeta\ in the 70–-80\% (80–-90\%) centrality class
about 3\% (10\%) lower than the value one would obtain with a centrality
selection based on the impact parameter.

For all the collisions in the 0--90\% centrality range the coordinates of the primary interaction point
can be reconstructed with good accuracy by correlating hits in the two
SPD layers. The measurement of the charged-particle multiplicity density
at mid-rapidity uses information from the SPD.
The acceptance of the SPD for charged particles spans different pseudorapidity regions
depending on the position of the interaction point along the beam line, $z$.
For example, for collisions with the vertex located within $\abs{z}<7\,\mathrm{cm}$ a maximum
acceptance of $\abs{\eta}<1.5$ can be reached, with approximately constant acceptance
for $\abs{\eta}<0.5$. To extend the pseudorapidity coverage up to $\abs{\eta}<2$,
all collisions with a primary vertex located within $\abs{z}<20\,\mathrm{cm}$ have been considered.


Following the method developed
earlier~\cite{Abbas:2013bpa,PhysRevLett.106.032301,Adam:2015ptt,Adam:2016ddh,Adam:2015kda},
tracklets (short track segments) are formed using the position of the primary
vertex and all possible combinations of hits between the two SPD layers.
The primary charged-particle multiplicity density \dndeta\ is obtained
from the number of tracklets that pass the quality selection criteria,
after correcting for detector acceptance,
reconstruction and selection efficiencies and contamination
from combinatorial background and secondary charged particles.
This selection allows primary charged-particle detection down to a momentum of \unit[30]{MeV/$c$}.
The corrections are estimated using a detailed simulation
based on events generated with the HIJING event generator \cite{Wang:1991hta} with particle transport in ALICE performed by GEANT3~\cite{GEANT3}.
The decay products of
long-lived decaying particles like $\mathrm{K}^{0}_{\mathrm{S}}$,
$\Lambda$, $\bar{\Lambda}$ and other strange hadrons are
classified as secondary particles \cite{ALICE-PUBLIC-2017-005} and
the contamination from these particles is subtracted from data.
It is known that HIJING underestimates
the relative production rate of strange particles in high-energy heavy-ion collisions.
For this reason, the simulation has been reweighed to reproduce
the relative particle abundances observed in the data which are about 30\% (50\%)
higher than HIJING in the most central (peripheral) collisions.
The reweighing factors have been derived from an estimate of
$\mathrm{K}^{0}_{\mathrm{S}}$, $\Lambda$ and $\bar{\Lambda}$ relative production
in the data, obtained via invariant mass reconstruction and compared to HIJING.

The deposited energy signal in the FMD is used to measure the
charged-particle pseudorapidity density in the forward regions
($-3.5<\eta<-1.8$ and $1.8<\eta<5$), following the method described
elsewhere~\cite{Abbas:2013bpa}.
The energy loss is measured in the 51,200 Si strip sensors of the detector and
a statistical approach is used to calculate the inclusive number
of charged particles.
A data-driven correction derived from previous studies~\cite{Adam:2015kda}
corrects for the background of secondary particles,
which are abundant in the forward regions.

\section{Systematic uncertainties}
\label{sec:sysuncer}

The systematic uncertainties on \avNpart\ are obtained by
varying the parameters of the Glauber model independently within their
estimated uncertainties and repeating the NBD-Glauber fit.
The uncertainty due to the centrality determination is estimated by
changing the value of V0 amplitude that corresponds to the top $90\%$
of the hadronic cross section by $\pm 0.5\%$.
This results in
an uncertainty on \dNdeta\ of 0.1\% to 4.8\% from central to
peripheral collisions. An additional 4\% uncertainty assigned to the
most peripheral class, arising from the remaining contamination from EM
processes, was estimated by studying the energy deposition in the ZDCs~\cite{PhysRevLett.109.252302}.

For the tracklet analysis at mid-rapidity the relative systematic
uncertainty on the measurement of the charged-particle multiplicity
in peripheral (central) events arises from the following sources:
tracklet selection 0.1\% (0.8\%), calculated by varying the tracklet
quality cut up to 4 times the nominal value;
combinatorial background
subtraction 0.5\% (2.0\%), estimated from simulations and
cross-checked using an alternative method where artificial SPD clusters are added to the data
and the number of corresponding artificial reconstructed tracklets are used for background subtraction;
particle composition 0.2\% (0.2\%), estimated by
changing the relative abundances of protons, pions and kaons by $\pm$30\%
in the simulation;
contamination by weak decays 0.3\% (0.3\%), estimated by changing
the reweighting factors;
extrapolation to zero transverse momentum 0.6\% (0.6\%), obtained from
the variation of the estimated yield of particles at low transverse
momentum by a factor of two in the simulation;
variations in detector acceptance and
efficiency 1\% (1\%), evaluated by carrying out the analysis for
different slices of the $z$-position of the interaction vertex
and with subsamples in azimuth.
At forward rapidities, the uncertainties related to the measurement of
multiplicity arise from the following sources:
the data-driven correction for secondary particles~\cite{Adam:2016ddh} 6.1\%;
the merging algorithms of signals from Si strips to a single particle 1\%;
variation in rejection threshold for calculation of the charged-particle
multiplicity per event $^{+1\%}_{-2\%}$;
particle composition 2\%, estimated in the same way as in the tracklet analysis.

The systematic uncertainties from centrality selection and
electromagnetic interactions affect the overall normalisation of the results.
The total systematic uncertainty, obtained by adding in quadrature all
contributions, amounts to 6.4\% (2\%) for peripheral (central) in
$|\eta|<2$, to 6.9\% for $\eta>3.5$ and to 6.4\% elsewhere in the forward
region, and is partially correlated over $\eta$ and between different
centrality classes.

\section{Results}
\label{sec:results}

\begin{table}[t]
\centering
\begin{tabular}{@{} c|c|c|c|c|c @{}} 
Centrality  & \avNpart & \dNdeta & \dNdetape & \Ntot &  $\frac{2}{\langle N_\mathrm{part}\rangle} N^\mathrm{tot}_\mathrm{ch}$ \\
\hline

0--1\%   & 246  $\pm$ 2 & 1302  $\pm$ 17		& 10.6 $\pm$ 0.2 & 14700  $\pm$ 300 & 119.5  $\pm$ 2.6\\
1--2\%   & 241  $\pm$ 2 & 1223  $\pm$ 25		& 10.1 $\pm$ 0.2 & 13840  $\pm$ 250 & 114.9  $\pm$ 2.3\\
2--3\%   & 236  $\pm$ 3 & 1166  $\pm$ 23		& 9.88 $\pm$ 0.23 &  13250  $\pm$ 280 & 112.3  $\pm$ 2.8\\
3--4\%   & 231  $\pm$ 2 & 1113  $\pm$ 20		& 9.64 $\pm$ 0.19 &  12700  $\pm$ 290 & 110.0  $\pm$ 2.7 \\
4--5\%   & 225  $\pm$ 3 & 1069  $\pm$ 20		& 9.50 $\pm$ 0.22 &  12180  $\pm$ 260 & 108.3  $\pm$ 2.7\\
\hline
0--2.5\%   & 242  $\pm$ 2 & 1238  $\pm$ 25		& 10.2 $\pm$ 0.2 & 14100  $\pm$ 320 & 116.5  $\pm$ 2.8\\
2.5--5.0\% & 229  $\pm$ 2 & 1096  $\pm$ 27 		& 9.57  $\pm$ 0.25 & 12440  $\pm$ 280 & 108.6  $\pm$ 2.6\\
5.0--7.5\% & 214  $\pm$ 3 &  986  $\pm$ 25 		& 9.21 	$\pm$ 0.27 & 11230  $\pm$ 330 & 105.0  $\pm$ 3.4\\
7.5--10\%  & 199  $\pm$ 2 &  891  $\pm$ 24 		& 8.95	$\pm$ 0.26 & 10300  $\pm$ 300 & 103.5  $\pm$ 3.2\\
\hline
0--5\%     & 236  $\pm$ 2    & 1167  $\pm$ 26	& 	9.89 	$\pm$ 0.24 & 13230  $\pm$ 280 & 112.1  $\pm$ 2.6  \\
5--10\%    & 207  $\pm$ 3    &  939  $\pm$ 24   &		9.07	$\pm$ 0.27 & 10820  $\pm$ 280 & 105.0  $\pm$ 3.1	\\
10--20\%   & 165  $\pm$ 3    &  706  $\pm$ 17   &		8.56  $\pm$ 0.26 & 8200  $\pm$ 310 & 99.4  $\pm$ 4.2 \\
20--30\%   & 118  $\pm$ 4    &  478  $\pm$ 11   &		8.10  $\pm$ 0.33 & 5670  $\pm$ 300 & 96.1  $\pm$ 6.0\\
30--40\%   &  82.2 $\pm$ 3.9 &  315  $\pm$ 8    &		7.66	$\pm$ 0.41 & 3770  $\pm$ 270 & 91.7  $\pm$ 7.9\\
40--50\%   &  54.6 $\pm$ 3.6 &  198  $\pm$ 5    & 	7.25	$\pm$ 0.51 & 2460  $\pm$ 220 & 90.1  $\pm$ 10\\
50--60\%   &  34.1 $\pm$ 3.0 &  118  $\pm$ 3    &		6.92	$\pm$ 0.63 & 1480  $\pm$ 170 & 86.8  $\pm$ 13\\
60--70\%   &  19.7 $\pm$ 2.1 &   64.7 $\pm$ 2.0 &		6.57	$\pm$ 0.73 & 828  $\pm$ 44 & 84.1  $\pm$ 10\\
70--80\%   &  10.5 $\pm$ 1.1 &   32.0 $\pm$ 1.3 &		6.10	$\pm$ 0.68 & 437  $\pm$ 16 & 83.2  $\pm$ 9.2\\
80--90\%   &  5.13  $\pm$ 0.46 &   13.3 $\pm$ 0.9 & 	5.19 $\pm$ 0.58 & 181  $\pm$ 7.0 & 70.6  $\pm$ 6.9 \\

\end{tabular}
\caption{\label{tab:data} The \dNdeta\ and \Ntot values for different centrality
  classes, defined by V0 multiplicity.  The errors are total
  uncertainties, the statistical contribution being negligible. The
  values of $\avNpart$ obtained with the Glauber model are also
  reported. The errors are obtained by varying the parameters of the
  NBD-Glauber calculation.}
\end{table}

\begin{figure}[t]
\begin{center}
\includegraphics[width=\figlen]{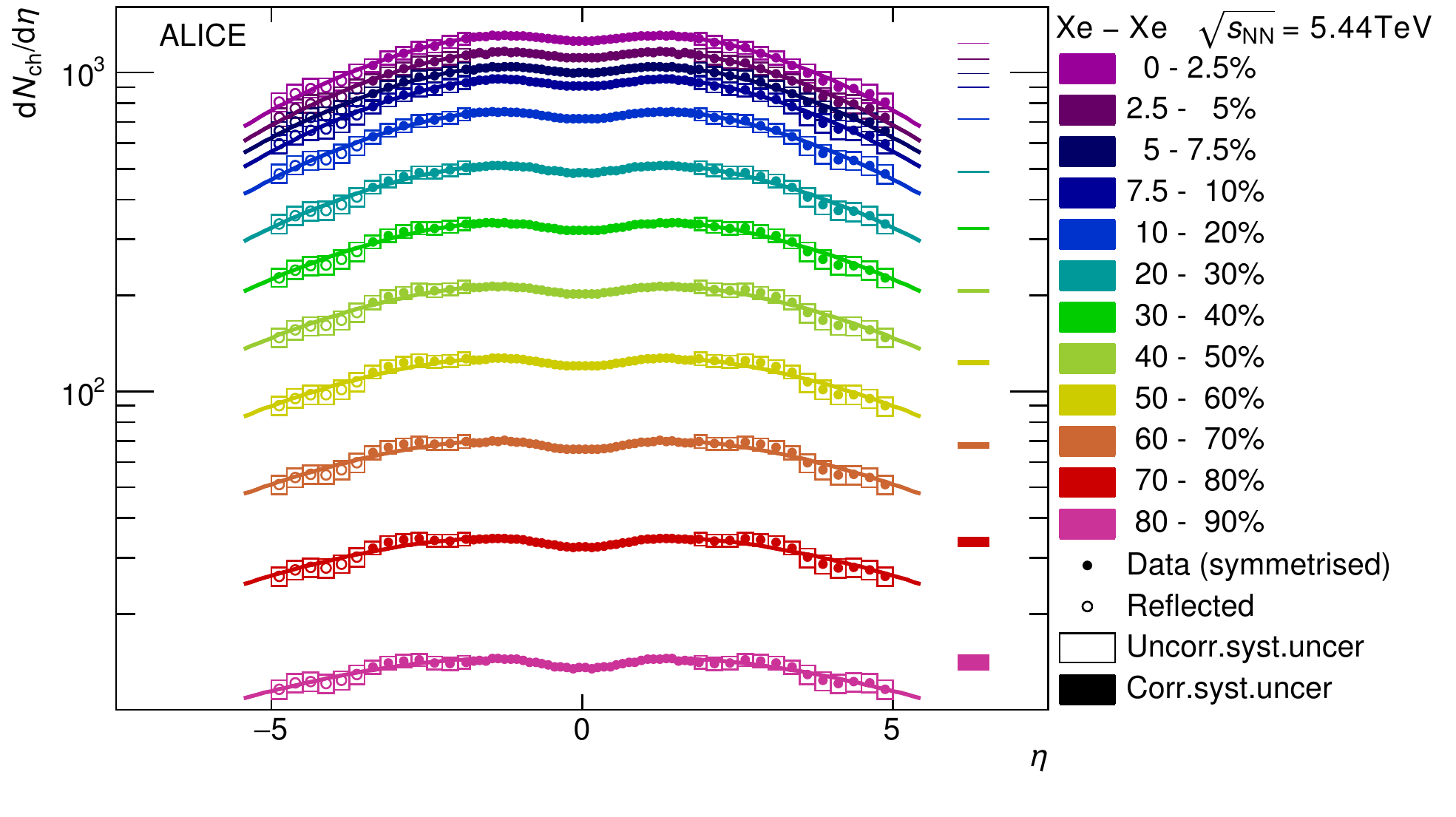}
\caption{ Charged--particle pseudorapidity density for
  12 centrality classes over a broad $\eta$ range in \XeXe\ collisions
  at \snn~=~\unit[5.44]{TeV}.  Boxes around the points reflect the total
  systematic uncertainties, while the filled squares on the right
  reflect the normalisation uncertainty from the centrality determination.
  Statistical errors are negligible.
  The reflection (open circles) of the $3.5<\eta<5$ values around
  $\eta=0$ is also shown.
  The lines correspond to fits to a gaussian distribution
  in rapidity multiplied by an effective Jacobian of transformation from
  $\eta$ to $y$.  }
\label{fig:dNdeta_05440}
\end{center}
\end{figure}

Figure~\ref{fig:dNdeta_05440} presents the charged-particle
multiplicity density \dndeta\ as a function of pseudorapidity
for 12 centrality classes.  The measurement is obtained from the SPD at
mid-rapidity, FMD in forward-rapidities, and combined in
regions of overlap ($1.8<|\eta|<2$) between the two detectors by
taking the weighted average using the non-shared uncertainties as
weights.  The data are symmetrised around $\eta=0$, averaging positive
and negative $\eta$ results wherever possible, and extended into the
non-measured region $-5 < \eta < -3.5$ by reflecting the $3.5 < \eta <
5$ values around $\eta=0$. Averaged values (left and right) agree within the uncertainties.
Assuming that the charged-particle rapidity density \dndy\ has Gaussian shape and using an
effective Jacobian, the measured \dndeta\ is fitted with this ansatz and a width of $\sigma=4.4\pm0.1$ is found, consistent with the value obtained in \PbPb\ at
\snn~=~\unit[5.02]{TeV}~\cite{Adam:2016ddh}.

The multiplicity density averaged over $|\eta|<0.5$ in different
centrality classes is shown in Tab.~\ref{tab:data}.
The total charged-particle multiplicity \Ntot\ is determined from the data in the
measured region and from extrapolations, up to $\eta = \pm y_{\mathrm{beam}}$, in the unmeasured region.
Three different functions are used to extrapolate the data points:
the difference of two Gaussian distributions centred at $\eta=0$; a Woods-Saxon-like
distribution in rapidity as proposed by PHOBOS~\cite{PhysRevC.83.024913}; and a
trapezoidal form.
The trapezoid ansatz in the forward
unmeasured regions corresponds to a linear extrapolation up to $\eta = \pm y_{\mathrm{beam}}$ with the
starting point constrained by the measurements.
A Gaussian \dndy\ in rapidity results in a distribution in pseudorapidity
which is very similar to the difference of two  Gaussians centered at $\eta$~=~0.
The central value in the
unmeasured regions ($-8.6<\eta<-3.5$ and $5<\eta<8.6$) is taken
as the average between the trapezoidal function (which gives the lowest $N^\mathrm{tot}_\mathrm{ch}$)
and the Gaussian \dndy\ (which gives the highest $N^\mathrm{tot}_\mathrm{ch}$).
The contribution from the extrapolated region is less than 30\% of $N^\mathrm{tot}_\mathrm{ch}$.
The systematic uncertainty of the extrapolated \Ntot\ is calculated as the
quadratic sum of contributions from the systematic uncertainty of the data
and a conservative contribution obtained by comparing the results from the
different fit functions. It amounts to about 14\% (4\%) of \Ntot\ in peripheral (central) events.
In order to compare bulk particle production at different energies and
in different collision systems, the average charged-particle multiplicity
density \dNdeta\ at mid-rapidity is divided by the average number of
participating nucleon pairs, \avNpart/2.
This allows one to compare nuclear collisions to \pp\ and \ppbar\ collisions.
The \avNpart\ values are
calculated within the Glauber model.

\begin{figure}[t]
\begin{center}
\includegraphics[width=\figlen]{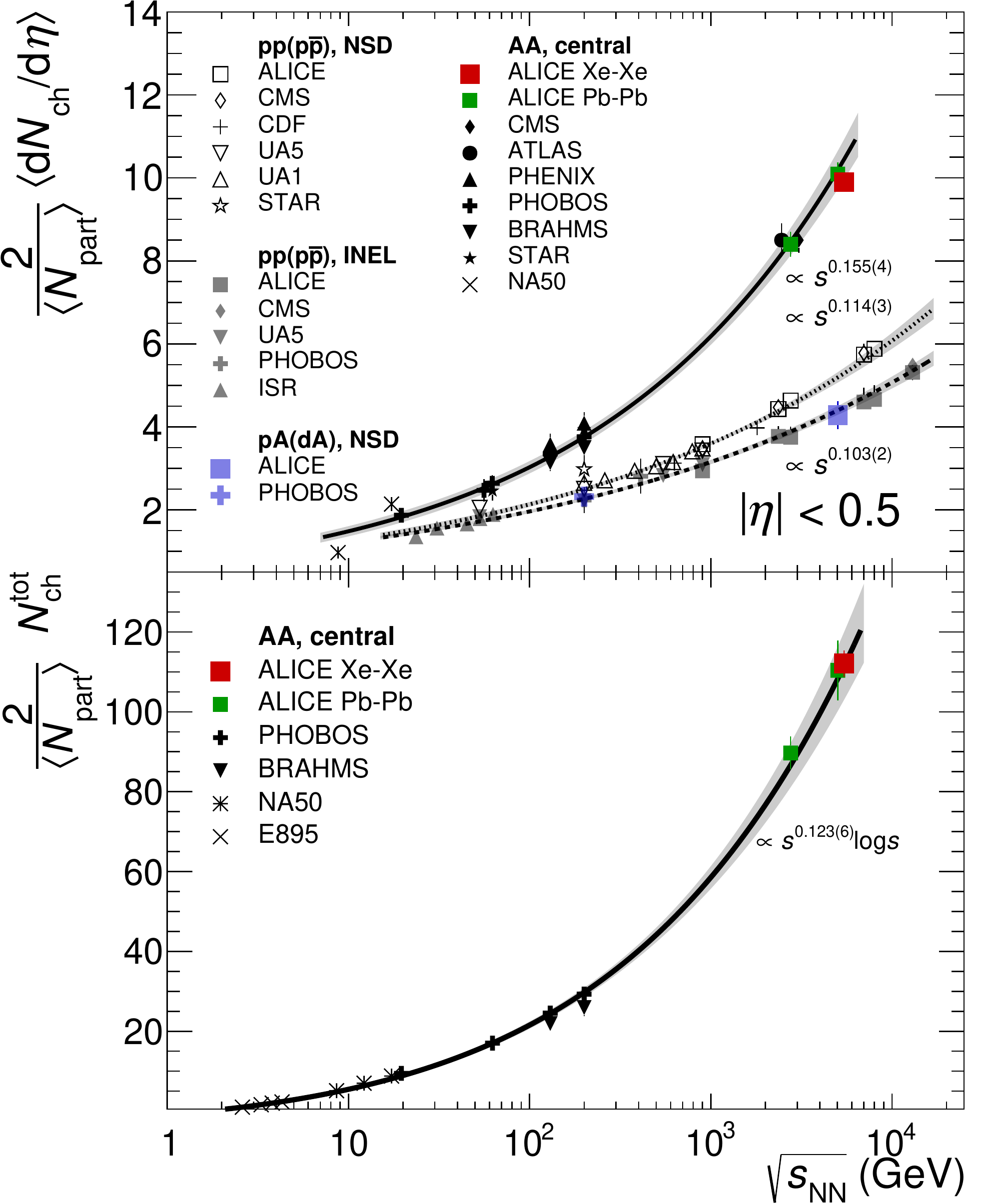}
\caption{Values of $\dNdetape$ (top) and $\frac{2}{\langle N_\mathrm{part}\rangle} N^\mathrm{tot}_\mathrm{ch}$ (bottom) for the 5\% most central
  \XeXe\ collisions compared to previous measurements in
  \PbPb\ \cite{Adam:2015ptt,Adam:2016ddh,PhysRevLett.106.032301,Aad2012363,Chatrchyan:2011aa,Abreu:2002fw}
  and
  \AuAu\ \cite{Bearden:2001xw,Bearden:2001qq,Adcox:2000sp,Alver:2010ck,Abelev:2008ab}
  as a function of \snn,
  as well as for inelastic \pp,
  \ppbar\ \cite{LongMultiPaper,Khachatryan2015143,Adam2016319} and
  non-single diffractive pA and dA collisions
  \cite{PhysRevLett.110.032301,Back:2003hx}.
  The lines are power law fits to the data, excluding \XeXe\ results.
	The central \PbPb\ measurements from CMS and ATLAS at \unit[2.76]{TeV} have been shifted
  horizontally for clarity.}
\label{fig:roots}
\end{center}
\end{figure}
Figure \ref{fig:roots} (top) shows the mid-rapidity charged-particle
multiplicity
normalised by the number of nucleon pairs participating in the collision,
\dNdetape, in \pp, $\mathrm{p\bar{p}}$, p(d)A
and in central heavy-ion collisions
as a function of the centre-of-mass
energy. The lines represent fits to lower energy results.
The \XeXe\ result is in agreement within the uncertainties
with the trend established from previous heavy-ion measurements,
which shows a stronger rise as a function of \snn\
than for pp and p--Pb collisions.
Figure \ref{fig:roots} (bottom) shows the total
charged-particle multiplicity per participant nucleon pair $\frac{2}{\langle N_\mathrm{part}\rangle} N^\mathrm{tot}_\mathrm{ch}$
, which follows the trend for central heavy-ion collisions.

\begin{figure}[t]
\begin{center}
\includegraphics[width=\figlen]{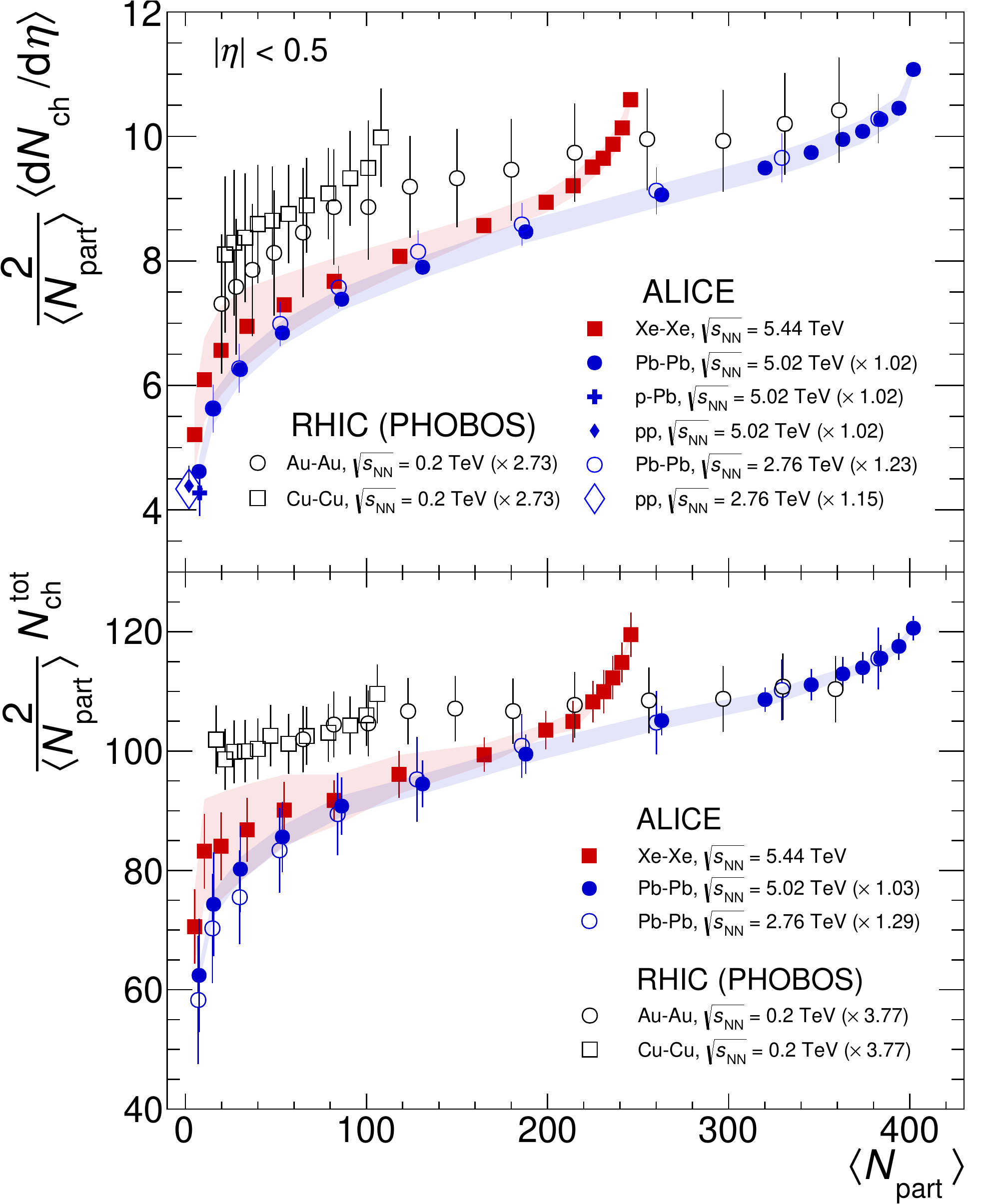}
\caption{The \dNdetape\ (top) and $\frac{2}{\langle N_\mathrm{part}\rangle} N^\mathrm{tot}_\mathrm{ch}$ (bottom) for
  \XeXe\ collisions at \snn~=~\unit[5.44]{TeV} as a function of \avNpart.
  The error bars indicate the point-to-point centrality-dependent
  uncertainties whereas the shaded band shows the correlated
  contributions. Also shown in the figure is the result from
  inelastic \pp\ at \s~=~\unit[5.02]{TeV} as well as non-single diffractive
  \pPb\ collisions~\cite{PhysRevLett.110.032301} and \PbPb\ collisions at \snn~=~\unit[5.02]{TeV}~\cite{Adam:2015ptt,Adam:2016ddh}.
  Note that \PbPb\ data at \snn~=~\unit[5.02]{TeV} were re-analysed in narrower centrality classes.
  Data from lower energies at \snn~=~\unit[2.76]{TeV}
  \cite{PhysRevLett.106.032301,Abbas:2013bpa} and \unit[200]{GeV}
  \cite{Alver:2010ck} are shown for comparison.}
\label{fig:npart}
\end{center}
\end{figure}

\begin{figure}[t]
\begin{center}
\includegraphics[width=\figlen]{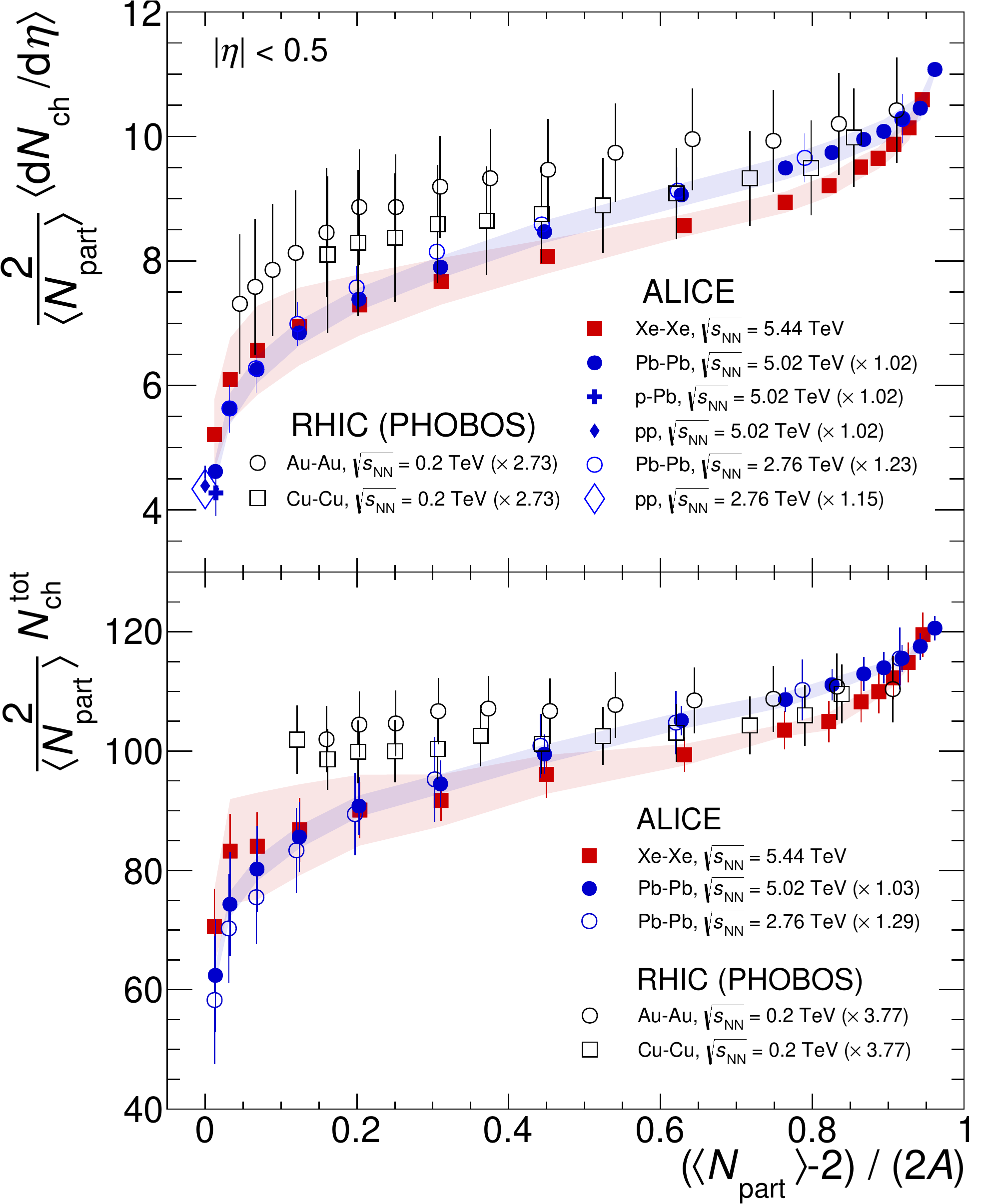}
\caption{The \dNdetape\ (top) and $\frac{2}{\langle N_\mathrm{part}\rangle} N^\mathrm{tot}_\mathrm{ch}$ (bottom) for
  \XeXe\ collisions at \snn~=~\unit[5.44]{TeV} as a function of $(\avNpart-2)/(2A)$.}
\label{fig:npartscaled}
\end{center}
\end{figure}

Figure \ref{fig:npart} shows the centrality dependence of the mid-rapidity and the total multiplicities per participant nucleon pairs.
The point-to-point
centrality-dependent uncertainties are indicated by error bars whereas
the shaded bands show the correlated uncertainties.
The values of \dNdetape\ and $\frac{2}{\langle N_\mathrm{part}\rangle} N^\mathrm{tot}_\mathrm{ch}$ decrease by a factor 2 from the most central to the most
peripheral collisions, where they agree with the values measured in minimum
bias \pp\ and \pPb\ collisions \cite{LongMultiPaper,PhysRevLett.110.032301}. The
data are compared to lower energy results at \snn~=~\unit[200]{GeV}
\cite{Alver:2010ck} for the RHIC experiment, \snn~=~\unit[2.76]{TeV}
\cite{PhysRevLett.106.032301,Abbas:2013bpa} and \snn~=~\unit[5.02]{TeV}~\cite{Adam:2015ptt,Adam:2016ddh}
for Pb--Pb collisions where the latter has been re-analysed with the same analysis
technique in narrower centrality classes, scaled to match the Xe--Xe data at \snn~=~\unit[5.44]{TeV}.
The scaling factors are calculated using the fit function of Fig.~\ref{fig:roots}
for the top 5\% central collisions.
For the 5\% most central Xe--Xe and for the 2\% most central Pb--Pb collisions, the \dNdetape\ increases steeply.
A similar conclusion was also reached for the RHIC
data \cite{Alver:2010ck}: the Cu--Cu trend resembles that
of Au--Au up to the most central
collisions and rises above it for the most central collisions. The
RHIC data are also shown in Fig.~\ref{fig:npart} and a deviation
from the LHC data for $\Npart<100$ is visible,
although with large uncertainties.
The steeper rise might be due to multiplicity fluctuations in the tail of
the Xe--Xe V0 amplitude distribution~\cite{ALICE-PUBLIC-2018-003}. The fluctuations occur both in the number of collisions over participants and in the number of charged
particles over participants. The rise is quantitatively reproduced by the
NBD-Glauber fit. The total number of charged particles scaled by the number of
participant pairs shows a slight increase as a function of the number of
participants in Fig.~\ref{fig:npart} (bottom), similar to that of the
midrapidity results, albeit with larger experimental uncertainties.
Figure~\ref{fig:npartscaled} shows the Xe--Xe and Pb--Pb results as a function
of a different scaling variable $(\avNpart-2)/(2A)$, where $A$ is the atomic
mass number of the colliding nucleus. 
The figure shows that $\dNdetape$ and $\frac{2}{\langle N_\mathrm{part}\rangle} N^\mathrm{tot}_\mathrm{ch}$ 
have a similar dependence on the number of participants 
relative to the possible maximum number of participants, which indicates a stronger dependence 
on geometric properties of the collision zone than on the collision system sizes. 


\begin{figure}[t]
\begin{center}
\includegraphics[width=\figlen]{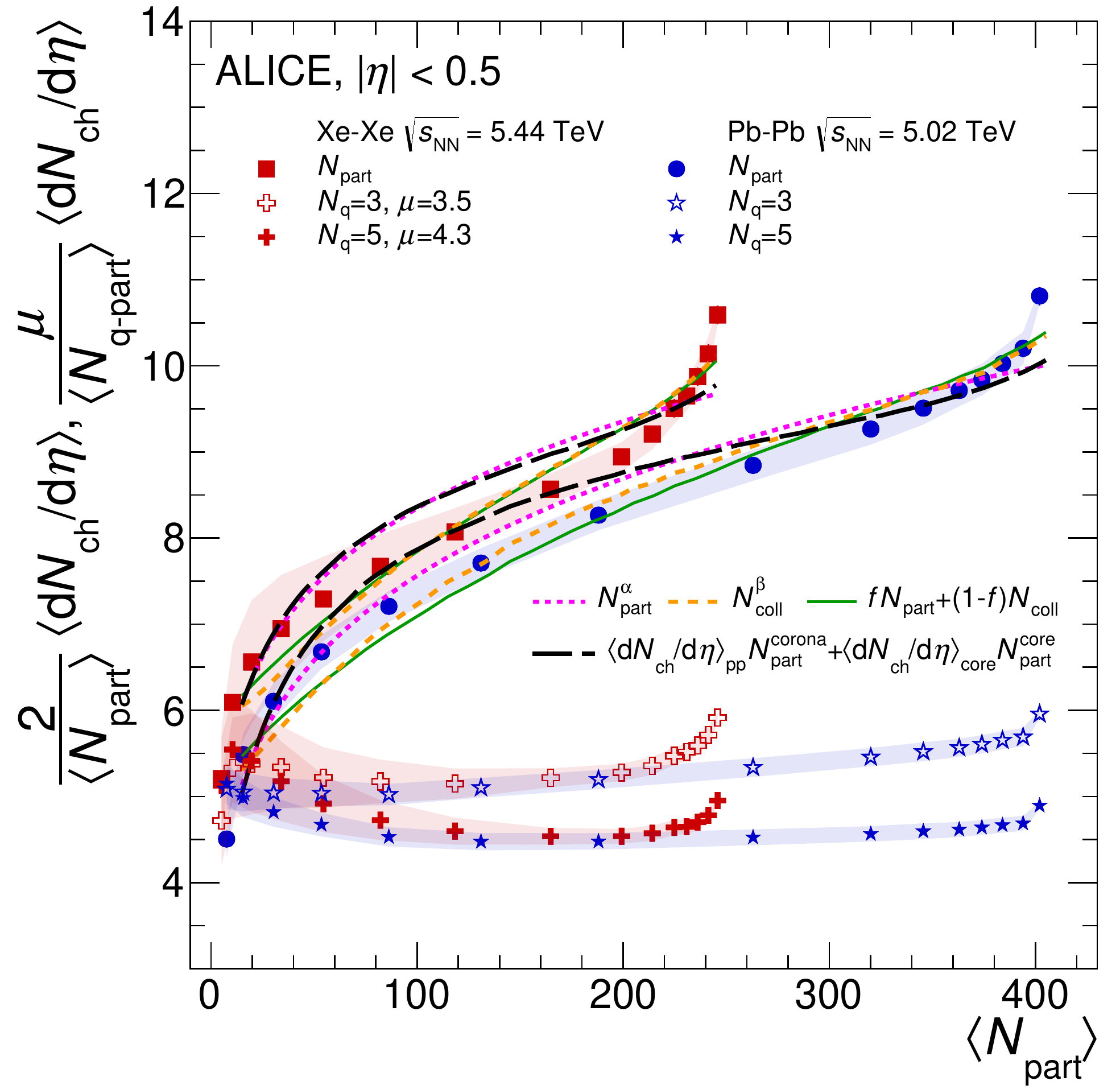}
\caption{The $\dNdetape$ for \XeXe\ collisions at \snn~=~\unit[5.44]{TeV} and
  \PbPb\ collisions at \snn~=~\unit[5.02]{TeV}~\cite{Adam:2015ptt}, as a function of $\langle
  N_{\mathrm{part}} \rangle$.  The \PbPb\ data are fitted with various
  parameterisations of \Npart\ and \Ncoll, calculated with the Glauber
  model.
  The same functions, with the values of the parameters from the Pb--Pb fit, are used for
  the \XeXe\ data.
	Also shown is \dNdeta\ per participant quark, \Ncpart,
  calculated with the effective wounded constituent quarks model~\cite{Loizides:2016djv}, as a
  function of \avNpart. The number of participant quarks \Ncpart\ is
  normalised by the average number of participant quarks in \pp\ collisions,
  $\mu$.}
\label{fig:qpart}
\end{center}
\end{figure}

The study of the centrality dependence
of particle multiplicity for different collision systems provides a variable number of
nucleon-nucleon collisions at equal number of participating nucleons and therefore
may provide further information to clarify the measured deviation
from \Npart\ scaling.
The scaling of the charged-particle multiplicity by the number of participant nucleons
was studied in detail and a deviation from \Npart-scaling was observed
at RHIC energies
\cite{Abreu:2002fw,Antinori:2001qn,Aggarwal:2001,Adare:2015bua,Alver:2010ck}.
The deviation from \Npart-scaling was initially thought to be due to a 
relative increase in hard processes in central collisions, but no conclusive evidence was found 
to support this interpretation.
Figure~\ref{fig:qpart} compares \dNdetape\ in \XeXe\ collisions
at \snn~=~\unit[5.44]{TeV} with different parameterisations for particle
production. Specifically, we used the two-component model
in Eq.~\ref{eqtwocomponent}
and two power-law functions $\dNdeta \propto \Npart^{\alpha}$
and $\dNdeta \propto \Ncoll^{\beta}$.
The functions were fitted to the \PbPb\ data at \snn~=~\unit[5.02]{TeV}~\cite{Adam:2015ptt}.
For the \XeXe\ data
only the absolute normalisation was adjusted.
The values of the parameters are also consistent with those obtained
at SPS and RHIC energies \cite{Adler2005,Abreu:2002fw}.
While no unique physics conclusion can
be drawn from such fits, this suggests that geometrical arguments may be
sufficient to provide a good description of particle production across
different colliding systems and beam energies.

Describing particle production in relativistic heavy-ion collisions
as a superposition of emission from a thermal core and hard scatterings
in a corona \cite{Werner:2007bf}, one can classify the participating nucleons
into those that scatter only once (\Npartcorona) and those that scatter
multiple times (\Npartcore).
The multiplicity can then be
fitted with the sum of those contributions, $\dNdeta_{\mathrm
  {pp}}\, \Npartcorona + \dNdeta_{\mathrm{core}}\, \Npartcore$,
where $\dNdeta_{\mathrm {pp}}$ is the multiplicity measured in
inelastic \pp\ collisions \cite{LongMultiPaper} and
$\dNdeta_{\mathrm{core}}$ is the contribution to the charged-particle
multiplicity from the core of the fireball, which is fitted to the data.
Figure~\ref{fig:qpart} also shows \dNdeta\ per participant quark
\Ncpart\ calculated with a Glauber model using effective wounded
constituent quarks~\cite{Eremin:2003qn}\cite{Loizides:2016djv}, as a function of \Npart, as was done for \PbPb\ collisions at
\snn~=~\unit[5.02]{TeV}~\cite{pubNote5023} that have been re-analysed in narrower centrality classes.
In the implementation of the quark-Glauber model the partonic degrees of freedom
(3 or 5) are located around the nucleon centres \cite{Loizides:2016djv}.
The effective inelastic scattering cross section for collisions of
constituent quarks is set to \unit[20.38]{mb} and \unit[9.76]{mb}, for $\Ncn=3$ and
$\Ncn=5$, respectively, adjusted to reproduce the \unit[68.4]{mb}
nucleon--nucleon inelastic cross section at \unit[5.44]{TeV}. \Ncpart\ has
been divided by the average value in pp collisions $\mu$ = \avNcpart,
which is 3.5 (4.3) for $\Ncn=3$ ($\Ncn=5$).
Comparing the behaviour of \dNdeta\ in terms of the dependence on the number of
nucleon or quark participants in the collision, one concludes that
\Ncpart\ scaling describes the data better than \Npart\ scaling
as previously observed \cite{pubNote5023,Adare:2015bua}
except the 0--10\% centrality range in Xe--Xe collisions
where a clear scaling violation is observed.

\begin{figure}[t]
\begin{center}
\includegraphics[width=\figlen]{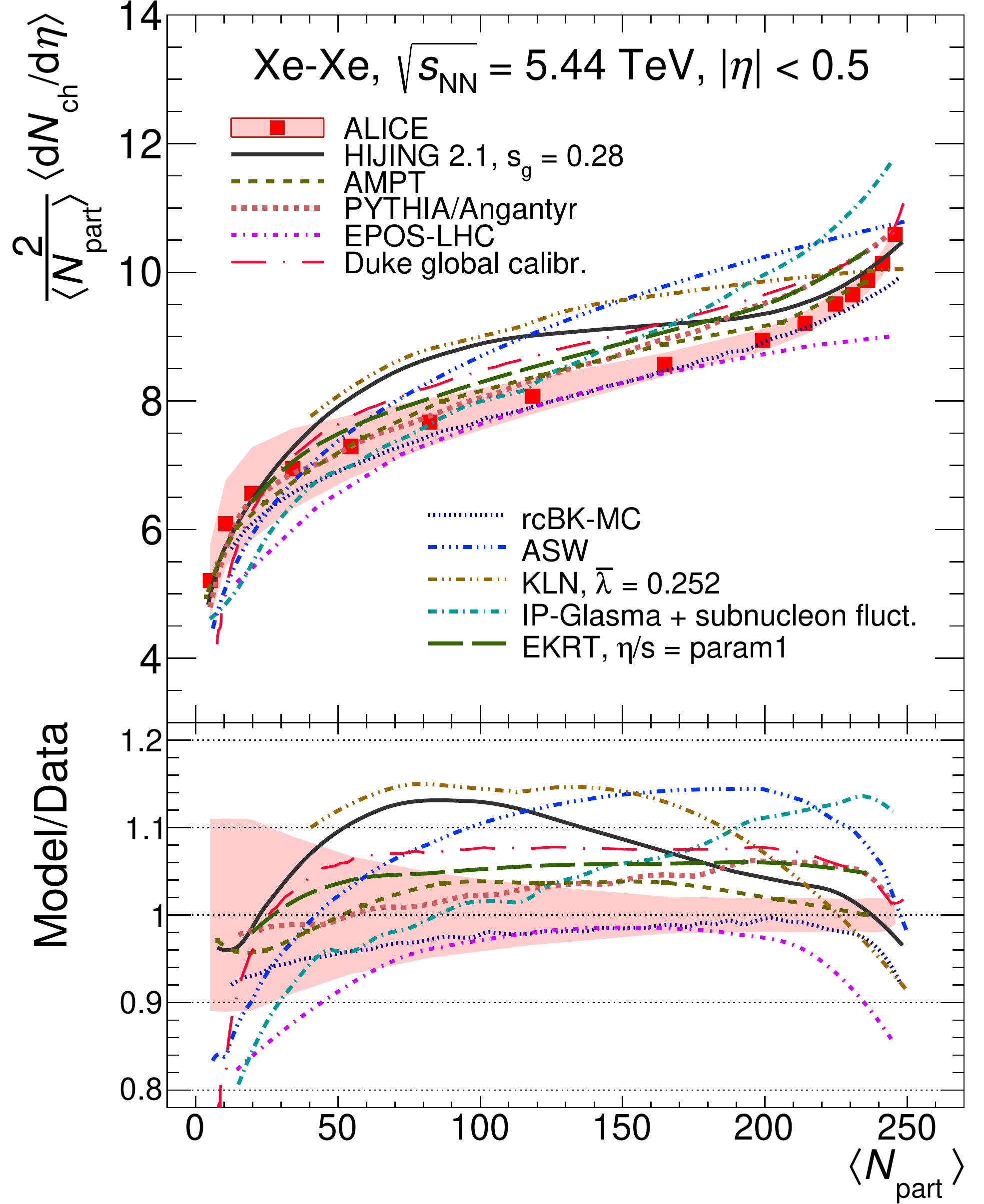}
\caption{The $\dNdetape$ for \XeXe\ collisions at \snn~=~\unit[5.44]{TeV} as a function of $\langle
  N_{\mathrm{part}} \rangle$ compared to
  model predictions
  \cite{Xu:2012au,PhysRevC.83.014915,PhysRevC.92.034906,Lin:2004en,Xu:2011fi,Bierlich:2016smv,Sjostrand:2014zea,Albacete:2012xq,Albacete:2010ad,PhysRevC.85.044920,Kharzeev:2001gp,Kharzeev:2004if,Kharzeev:2000ph,Armesto:2004ud,Schenke:2013dpa,Schenke:2012wb,Niemi:2015qia,Niemi:2015voa,Eskola:2017bup}. The bottom panel shows the ratio of the models to the data.
  The shaded band around the points reflects the correlated systematic uncertainties.}
\label{fig:models}
\end{center}
\end{figure}

Figure \ref{fig:models} shows a comparison of the \XeXe\ data to
calculations from theoretical models at mid-rapidity.
HIJING 2.1~\cite{Xu:2012au,PhysRevC.83.014915}
combines perturbative QCD processes with soft interactions,
and includes a strong impact parameter dependence of
parton shadowing~\cite{ASHMAN1988603}.
For \XeXe\ data at \snn~=~\unit[5.44]{TeV} it uses a large gluon
shadowing parameter of 0.28 to limit the multiplicity per participant.
With this choice, the same as in \PbPb\ collisions at \snn~=~\unit[5.02]{TeV},
the multiplicities at mid-rapidity and the centrality dependence in the most central collisions are reproduced.
AMPT~\cite{Lin:2004en,Xu:2011fi} is a model which implements hydrodynamical
evolution of an initial state produced by HIJING. It includes spatial
coalescence of quarks to hadrons, followed by hadronic scattering.
AMPT describes both the shape and the overall magnitude of the mid-rapidity data.
PYTHIA/Angantyr~\cite{Bierlich:2016smv} extends the nucleon--nucleon
model of PYTHIA 8.230~\cite{Sjostrand:2014zea} to the case of
heavy-ion collisions, essentially performing individual
nucleon--nucleon collisions at the parton level,
while the resulting Lund-strings are hadronised as an ensemble.
It is interesting to note that this model
agrees reasonably well with the data even though it was developed as an
extension of a generator for nucleon--nucleon collisions.
EPOS LHC~\cite{PhysRevC.92.034906} is a parton model based on the
Gribov-Regge theory,
designed for minimum bias hadronic interactions, which incorporates
collective effects treated via a flow parameterisation and a separation
of the initial state into core--corona parts.
The shape of the centrality dependence is reproduced fairly well
at intermediate centralities, however,
the model underestimates the absolute values of the
multiplicity, as was the case in \PbPb\ collisions at \snn~=~\unit[5.02]{TeV}
\cite{Adam:2015ptt}.
The Duke global calibrated model is based on a Bayesian Statistics analysis
using T\raisebox{-.5ex}{R}ENTo initial conditions for high-energy nuclear
collisions~\cite{Bass:2017zyn, Moreland:2014oya}. The subsequent transport
dynamics is then simulated using the iEBE-VISHNU event-by-event simulations
for relativistic heavy-ion collisions which uses a hybrid approach based on
(2 + 1)-dimensional viscous hydrodynamics coupled to a hadronic cascade model~\cite{Shen:2014vra}.
The Duke global calibrated model can reproduce the shape of the mid-rapidity distribution, but overestimates slightly the overall magnitude.

\begin{figure}[t]
\begin{center}
\includegraphics[width=\figlen]{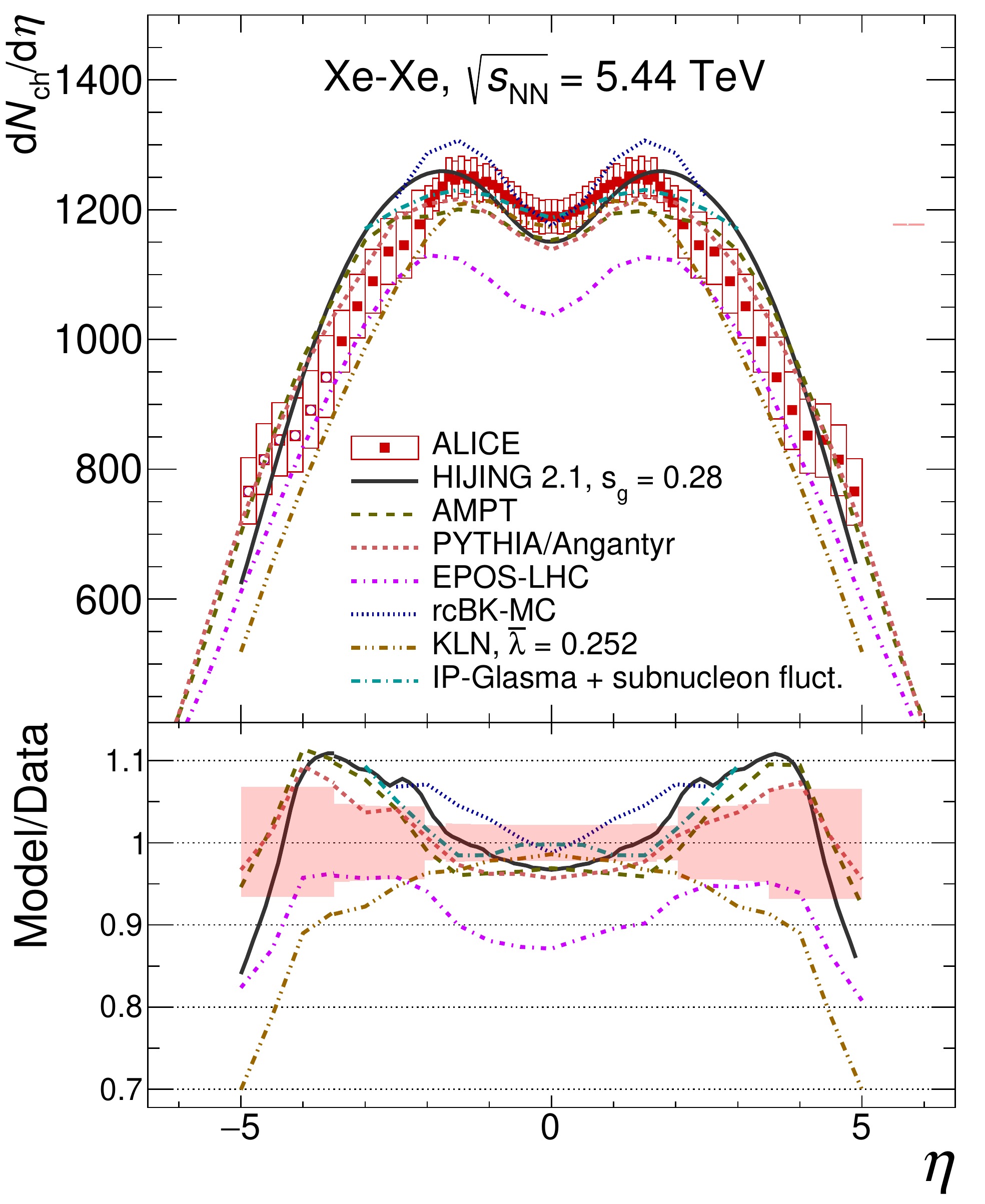}
\caption{Comparison of \dndeta\ as a function of $\eta$ in the 0--5\%
  central class to model predictions. The bottom panel shows the ratio
  of the models to the data. Boxes around the points reflect the total
  uncorrelated systematic uncertainties.}
\label{fig:Models3_05440}
\end{center}
\end{figure}

Saturation-inspired models
(rcBK-MC \cite{Albacete:2012xq,Albacete:2010ad},
KLN \cite{PhysRevC.85.044920,Kharzeev:2001gp,Kharzeev:2004if,Kharzeev:2000ph},
ASW \cite{Armesto:2004ud},
IP-Glasma \cite{Schenke:2013dpa,Schenke:2012wb}
and EKRT \cite{Niemi:2015qia,Niemi:2015voa,Eskola:2017bup})
rely on perturbative QCD and an energy-dependent
saturation scale, which
limits the number of produced partons, and in turn the number of
produced particles.
This results in a factorisation of the energy and centrality
dependence of particle production or, in other words, in the invariance
of the centrality growth, as observed in the
experimental data~\cite{PhysRevC.70.021902}.
The rcBK-MC model limits the centrality growth using the rc-BK equation.
It provides a good description of the mid-rapidity data,
both of the shape and the highest multiplicity reached in central
collisions.  The ASW prediction overestimates the data, while it was very
accurate in \PbPb\ at \snn~=~\unit[5.02]{TeV}. The KLN model does not
describe the shape well and, although it agrees with the value
measured for most central collisions, it is significantly above the
centrality dependence of the data.
The IP-Glasma model naturally produces initial energy fluctuations
computed within the Color Glass Condensate framework combining an
impact parameter dependent saturation model.
It uses a gluon multiplicity scaled to describe hadron multiplicities
measured in \PbPb\ collisions at \snn~=~\unit[5.02]{TeV}~\cite{Adam:2015ptt}.
The centrality dependence is stronger than that observed in mid-rapidity data
over(under)-predicting the data in central (peripheral) collisions.
The EKRT model for heavy-ion collisions uses
perturbative QCD with a conjecture of gluon saturation to suppress
soft parton production. The saturation scale is also dependent
on the local product of thickness functions, implying a geometrical scaling.
The space-time evolution of the system is then described with
viscous fluid dynamics event-by-event. The normalisation is fixed by
exploiting the 0--5\% most central multiplicity measurement in
\PbPb\ collisions at \snn~=~\unit[2.76]{TeV}~\cite{Aamodt:2010pb}. As for
\PbPb\ collisions at \snn~=~\unit[5.02]{TeV}, the EKRT model can describe both
the shape and the overall magnitude of multiplicity
on centrality.
In general, almost all models reproduce the steep rise versus \avNpart\ while
EPOS-LHC, ASW and KLN show a saturation behaviour. The predictions show
a similar trend as for the Pb--Pb case~\cite{Adam:2015ptt} and altogether a flatter distribution
with respect to data.

In Fig.~\ref{fig:Models3_05440}, the models are compared to the
pseudorapidity dependence of the \dndeta\ for the top 5\% central collisions.
HIJING 2.1 reproduces the pseudorapidity dependence at mid-rapidity well,
but overestimates the data at forward rapidity, due to the large value of
the shadowing parameter used.  AMPT and PYTHIA/Angantyr describe the
data fairly well, with a slight overestimate at forward rapidities.
EPOS LHC reproduces the shape well, but under-predicts the multiplicity
overall.  The rcBK-MC is restricted to $|\eta| < 2.5$ since its
formalism can only be used for rapidities far from the fragmentation regions.
It shows a narrower distribution than what is
seen in data. KLN agrees with the data at mid-rapidity, but not
at forward rapidity, where it under-predicts the data. For IP-Glasma the
rapidity dependence is provided by the IP-Sat model~\cite{Rezaeian:2012ji} and it is converted to
pseudorapidity using an effective mass of \unit[0.2]{GeV}/$\mathrm{c}^{2}$. The shape is wider
than that of the data. Regarding the case of the pseudorapidity dependence all the models
show similar trends as for Pb--Pb collisions~\cite{Adam:2016ddh} except HIJING 2.1
which describes the Xe--Xe measurements better than the Pb--Pb data.

\section{Conclusions}
\label{sec:concl}

The measurements of the charged-particle multiplicity density
and its centrality dependence in \XeXe\ collisions at
\snn~=~\unit[5.44]{TeV} have been presented over the pseudorapidity range $-3.5<\eta<5$
using the full acceptance of the ALICE detector.
For the 5\% most central collisions, the average charged-particle pseudorapidity density at mid-rapidity ($|\eta|
< 0.5$) is $1167 \pm 26$ and the total number of charged particles is
$13230 \pm 280$.
Scaled by the number of participant pairs, these are found to
follow the same power-law dependence with energy established in previous heavy-ion
measurements.

The centrality dependences of \dNdetape\ and $\frac{2}{\langle
N_\mathrm{part}\rangle} N^\mathrm{tot}_\mathrm{ch}$ are very similar to those
previously measured in \PbPb\ collisions at similar or lower energies up to the
5\% most central \XeXe\ collisions, where the \XeXe\ results are larger than
the \PbPb\ results at a similar number of participating nucleons. Similar
conclusions were drawn at RHIC from the comparison of the data for Cu--Cu and
Au--Au collisions~\cite{PhysRevLett.102.142301}.
The steeper rise might be due to multiplicity fluctuations in the tail of the Xe--Xe V0 amplitude.

While measurements of particle production in large and medium-sized
colliding systems such as \XeXe\
are abundant and become even more precise, the underlying mechanism to
describe the increase with energy and centrality is still not completely
understood. Deeper insight of the system size dependence of
particle production may come from the study of light-nuclei collisions,
still not much explored at high energy, which could bridge the
gap between the trends observed in \pp\ and pA collisions and those of
the mid-sized \XeXe\ and the large \PbPb\ systems.

%
%

\newenvironment{acknowledgement}{\relax}{\relax}
\begin{acknowledgement}
\section*{Acknowledgements}
The ALICE Collaboration would like to thank N. Armesto, W.-T. Deng, A. Dumitru, K. Eskola, G. Levin, L. Lonnblad, S. Moreland, H. Niemi, T. Pierog and B. Schenke for helpful discussions on their model predictions.
 

The ALICE Collaboration would like to thank all its engineers and technicians for their invaluable contributions to the construction of the experiment and the CERN accelerator teams for the outstanding performance of the LHC complex.
The ALICE Collaboration gratefully acknowledges the resources and support provided by all Grid centres and the Worldwide LHC Computing Grid (WLCG) collaboration.
The ALICE Collaboration acknowledges the following funding agencies for their support in building and running the ALICE detector:
A. I. Alikhanyan National Science Laboratory (Yerevan Physics Institute) Foundation (ANSL), State Committee of Science and World Federation of Scientists (WFS), Armenia;
Austrian Academy of Sciences and Nationalstiftung f\"{u}r Forschung, Technologie und Entwicklung, Austria;
Ministry of Communications and High Technologies, National Nuclear Research Center, Azerbaijan;
Conselho Nacional de Desenvolvimento Cient\'{\i}fico e Tecnol\'{o}gico (CNPq), Universidade Federal do Rio Grande do Sul (UFRGS), Financiadora de Estudos e Projetos (Finep) and Funda\c{c}\~{a}o de Amparo \`{a} Pesquisa do Estado de S\~{a}o Paulo (FAPESP), Brazil;
Ministry of Science \& Technology of China (MSTC), National Natural Science Foundation of China (NSFC) and Ministry of Education of China (MOEC) , China;
Ministry of Science and Education, Croatia;
Ministry of Education, Youth and Sports of the Czech Republic, Czech Republic;
The Danish Council for Independent Research | Natural Sciences, the Carlsberg Foundation and Danish National Research Foundation (DNRF), Denmark;
Helsinki Institute of Physics (HIP), Finland;
Commissariat \`{a} l'Energie Atomique (CEA) and Institut National de Physique Nucl\'{e}aire et de Physique des Particules (IN2P3) and Centre National de la Recherche Scientifique (CNRS), France;
Bundesministerium f\"{u}r Bildung, Wissenschaft, Forschung und Technologie (BMBF) and GSI Helmholtzzentrum f\"{u}r Schwerionenforschung GmbH, Germany;
General Secretariat for Research and Technology, Ministry of Education, Research and Religions, Greece;
National Research, Development and Innovation Office, Hungary;
Department of Atomic Energy Government of India (DAE), Department of Science and Technology, Government of India (DST), University Grants Commission, Government of India (UGC) and Council of Scientific and Industrial Research (CSIR), India;
Indonesian Institute of Science, Indonesia;
Centro Fermi - Museo Storico della Fisica e Centro Studi e Ricerche Enrico Fermi and Istituto Nazionale di Fisica Nucleare (INFN), Italy;
Institute for Innovative Science and Technology , Nagasaki Institute of Applied Science (IIST), Japan Society for the Promotion of Science (JSPS) KAKENHI and Japanese Ministry of Education, Culture, Sports, Science and Technology (MEXT), Japan;
Consejo Nacional de Ciencia (CONACYT) y Tecnolog\'{i}a, through Fondo de Cooperaci\'{o}n Internacional en Ciencia y Tecnolog\'{i}a (FONCICYT) and Direcci\'{o}n General de Asuntos del Personal Academico (DGAPA), Mexico;
Nederlandse Organisatie voor Wetenschappelijk Onderzoek (NWO), Netherlands;
The Research Council of Norway, Norway;
Commission on Science and Technology for Sustainable Development in the South (COMSATS), Pakistan;
Pontificia Universidad Cat\'{o}lica del Per\'{u}, Peru;
Ministry of Science and Higher Education and National Science Centre, Poland;
Korea Institute of Science and Technology Information and National Research Foundation of Korea (NRF), Republic of Korea;
Ministry of Education and Scientific Research, Institute of Atomic Physics and Romanian National Agency for Science, Technology and Innovation, Romania;
Joint Institute for Nuclear Research (JINR), Ministry of Education and Science of the Russian Federation and National Research Centre Kurchatov Institute, Russia;
Ministry of Education, Science, Research and Sport of the Slovak Republic, Slovakia;
National Research Foundation of South Africa, South Africa;
Centro de Aplicaciones Tecnol\'{o}gicas y Desarrollo Nuclear (CEADEN), Cubaenerg\'{\i}a, Cuba and Centro de Investigaciones Energ\'{e}ticas, Medioambientales y Tecnol\'{o}gicas (CIEMAT), Spain;
Swedish Research Council (VR) and Knut \& Alice Wallenberg Foundation (KAW), Sweden;
European Organization for Nuclear Research, Switzerland;
National Science and Technology Development Agency (NSDTA), Suranaree University of Technology (SUT) and Office of the Higher Education Commission under NRU project of Thailand, Thailand;
Turkish Atomic Energy Agency (TAEK), Turkey;
National Academy of  Sciences of Ukraine, Ukraine;
Science and Technology Facilities Council (STFC), United Kingdom;
National Science Foundation of the United States of America (NSF) and United States Department of Energy, Office of Nuclear Physics (DOE NP), United States of America.    
\end{acknowledgement}

\bibliographystyle{utphys}   
\bibliography{dNdetaXe}

\providecommand{\href}[2]{#2}\begingroup\raggedright\begin{thebibliography}{10}

\bibitem{Karsch:2001cy}
F.~Karsch, ``Lattice {QCD} at high temperature and density,''
  \href{http://dx.doi.org/10.1007/3-540-45792-5_6}{{\em Lect. Notes Phys.}
  {\bfseries 583} (2002) 209--249},
\href{http://arxiv.org/abs/hep-lat/0106019}{{\ttfamily arXiv:hep-lat/0106019
  [hep-lat]}}.

\bibitem{Muller:2012zq}
B.~Muller, J.~Schukraft, and B.~Wyslouch, ``First results from {Pb+Pb}
  collisions at the {LHC},''
  \href{http://dx.doi.org/10.1146/annurev-nucl-102711-094910}{{\em Ann. Rev.
  Nucl. Part. Sci.} {\bfseries 62} (2012) 361--386},
\href{http://arxiv.org/abs/1202.3233}{{\ttfamily arXiv:1202.3233 [hep-ex]}}.

\bibitem{Alver:2010ck}
{\bfseries PHOBOS} Collaboration, B.~Alver {\em et~al.}, ``{PHOBOS results on
  charged particle multiplicity and pseudorapidity distributions in Au+Au,
  Cu+Cu, d+Au, and p+p collisions at ultra-relativistic energies},''
  \href{http://dx.doi.org/10.1103/PhysRevC.83.024913}{{\em Phys. Rev.}
  {\bfseries C83} (2011) 024913},
\href{http://arxiv.org/abs/1011.1940}{{\ttfamily arXiv:1011.1940 [nucl-ex]}}.

\bibitem{PhysRevLett.106.032301}
{\bfseries ALICE} Collaboration, K.~Aamodt {\em et~al.}, ``{Centrality
  dependence of the charged-particle multiplicity density at mid-rapidity in
  Pb-Pb collisions at $\sqrt{s_\mathrm{NN}}=2.76$ TeV},''
  \href{http://dx.doi.org/10.1103/PhysRevLett.106.032301}{{\em Phys. Rev.
  Lett.} {\bfseries 106} (2011) 032301},
\href{http://arxiv.org/abs/1012.1657}{{\ttfamily arXiv:1012.1657 [nucl-ex]}}.

\bibitem{Abbas:2013bpa}
{\bfseries ALICE} Collaboration, E.~Abbas {\em et~al.}, ``{Centrality
  dependence of the pseudorapidity density distribution for charged particles
  in \PbPb\ collisions at \snn\ = 2.76 TeV},''
  \href{http://dx.doi.org/10.1016/j.physletb.2013.09.022}{{\em Phys. Lett.}
  {\bfseries B726} (2013) 610--622},
\href{http://arxiv.org/abs/1304.0347}{{\ttfamily arXiv:1304.0347 [nucl-ex]}}.

\bibitem{Aad2012363}
{\bfseries ATLAS} Collaboration, G.~Aad {\em et~al.}, ``{Measurement of the
  centrality dependence of the charged particle pseudorapidity distribution in
  lead-lead collisions at $\sqrt{s_\mathrm{NN}}=2.76$ TeV with the ATLAS
  detector},'' \href{http://dx.doi.org/10.1016/j.physletb.2012.02.045}{{\em
  Phys. Lett.} {\bfseries B710} (2012) 363--382},
\href{http://arxiv.org/abs/1108.6027}{{\ttfamily arXiv:1108.6027 [hep-ex]}}.

\bibitem{Chatrchyan:2011aa}
{\bfseries CMS} Collaboration, S.~Chatrchyan {\em et~al.}, ``Dependence on
  pseudorapidity and on centrality of charged hadron production in {PbPb}
  collisions at \snn\ = 2.76 {TeV},''
  \href{http://dx.doi.org/10.1007/JHEP08(2011)141}{{\em JHEP} {\bfseries 08}
  (2011) 141},
\href{http://arxiv.org/abs/1107.4800}{{\ttfamily arXiv:1107.4800 [nucl-ex]}}.

\bibitem{Adam:2015ptt}
{\bfseries ALICE} Collaboration, J.~Adam {\em et~al.}, ``{Centrality dependence
  of the charged-particle multiplicity density at midrapidity in Pb-Pb
  collisions at \snn\ = 5.02 TeV},''
  \href{http://dx.doi.org/10.1103/PhysRevLett.116.222302}{{\em Phys. Rev.
  Lett.} {\bfseries 116} (2016) 222302},
\href{http://arxiv.org/abs/1512.06104}{{\ttfamily arXiv:1512.06104 [nucl-ex]}}.

\bibitem{Adam:2016ddh}
{\bfseries ALICE} Collaboration, J.~Adam {\em et~al.}, ``{Centrality dependence
  of the pseudorapidity density distribution for charged particles in Pb-Pb
  collisions at \snn\ = 5.02 TeV},''
  \href{http://dx.doi.org/10.1016/j.physletb.2017.07.017}{{\em Phys. Lett.}
  {\bfseries B772} (2017) 567--577},
\href{http://arxiv.org/abs/1612.08966}{{\ttfamily arXiv:1612.08966 [nucl-ex]}}.

\bibitem{LongMultiPaper}
{\bfseries ALICE} Collaboration, J.~Adam {\em et~al.}, ``{Charged-particle
  multiplicities in proton--proton collisions at $\sqrt{s} = 0.9$ to 8 TeV},''
  \href{http://dx.doi.org/10.1140/epjc/s10052-016-4571-1}{{\em Eur. Phys. J.}
  {\bfseries C77} no.~1, (2017) 33},
\href{http://arxiv.org/abs/1509.07541}{{\ttfamily arXiv:1509.07541 [nucl-ex]}}.

\bibitem{PhysRevLett.110.032301}
{\bfseries ALICE} Collaboration, B.~Abelev {\em et~al.}, ``Pseudorapidity
  density of charged particles in {p + Pb} collisions at
  $\sqrt{{s}_\mathrm{NN}}$ = 5.02 tev,''
  \href{http://dx.doi.org/10.1103/PhysRevLett.110.032301}{{\em Phys. Rev.
  Lett.} {\bfseries 110} (2013) 032301}.
  \url{http://link.aps.org/doi/10.1103/PhysRevLett.110.032301}.

\bibitem{Abelev:2014ffa}
{\bfseries ALICE} Collaboration, B.~Abelev {\em et~al.}, ``{Performance of the
  ALICE Experiment at the CERN LHC},''
  \href{http://dx.doi.org/10.1142/S0217751X14300440}{{\em Int. J. Mod. Phys.}
  {\bfseries A29} (2014) 1430044},
\href{http://arxiv.org/abs/1402.4476}{{\ttfamily arXiv:1402.4476 [nucl-ex]}}.

\bibitem{1748-0221-4-03-P03023}
R.~Santoro {\em et~al.}, ``{The ALICE Silicon Pixel Detector: Readiness for the
  first proton beam},''
\href{http://dx.doi.org/10.1088/1748-0221/4/03/P03023}{{\em JINST} {\bfseries
  4} (2009) P03023}.

\bibitem{Christensen:2007yc}
C.~H. Christensen, J.~J. Gaardhoje, K.~Gulbrandsen, B.~S. Nielsen, and
  C.~Sogaard, ``{The ALICE Forward Multiplicity Detector},''
  \href{http://dx.doi.org/10.1142/S0218301307008057}{{\em Int. J. Mod. Phys.}
  {\bfseries E16} (2007) 2432--2437},
\href{http://arxiv.org/abs/0712.1117}{{\ttfamily arXiv:0712.1117 [nucl-ex]}}.

\bibitem{Cortese:781854}
{\bfseries ALICE} Collaboration, P.~Cortese {\em et~al.}, {\em {ALICE forward
  detectors: FMD, TO and VO: Technical Design Report}}.
\newblock Technical Design Report ALICE. CERN, Geneva, 2004.
\newblock \url{https://cds.cern.ch/record/781854}.
\newblock Submitted on 10 Sep 2004.

\bibitem{Abbas:2013taa}
{\bfseries ALICE} Collaboration, E.~Abbas {\em et~al.}, ``{Performance of the
  ALICE VZERO system},''
  \href{http://dx.doi.org/10.1088/1748-0221/8/10/P10016}{{\em JINST} {\bfseries
  8} (2013) P10016},
\href{http://arxiv.org/abs/1306.3130}{{\ttfamily arXiv:1306.3130 [nucl-ex]}}.

\bibitem{PUDDU2007397}
E.~Puddu {\em et~al.}, ``The zero degree calorimeters for the {ALICE}
  experiment,''
  \href{http://dx.doi.org/https://doi.org/10.1016/j.nima.2007.08.013}{{\em
  Nuclear Instruments and Methods in Physics Research Section A: Accelerators,
  Spectrometers, Detectors and Associated Equipment} {\bfseries 581} no.~1,
  (2007) 397 -- 401}. VCI 2007.

\bibitem{Alver:2008aq}
B.~Alver, M.~Baker, C.~Loizides, and P.~Steinberg, ``{The PHOBOS Glauber Monte
  Carlo},''
\href{http://arxiv.org/abs/0805.4411}{{\ttfamily arXiv:0805.4411 [nucl-ex]}}.

\bibitem{Loizides:2014vua}
C.~Loizides, J.~Nagle, and P.~Steinberg, ``{Improved version of the PHOBOS
  Glauber Monte Carlo},''
  \href{http://dx.doi.org/10.1016/j.softx.2015.05.001}{{\em SoftwareX}
  {\bfseries 1-2} (2015) 13--18},
\href{http://arxiv.org/abs/1408.2549}{{\ttfamily arXiv:1408.2549 [nucl-ex]}}.

\bibitem{PhysRevC.88.044909}
{\bfseries ALICE} Collaboration, B.~Abelev {\em et~al.}, ``{Centrality
  determination of Pb-Pb collisions at $\sqrt{s_\mathrm{NN}}$ = 2.76 TeV with
  ALICE},'' \href{http://dx.doi.org/10.1103/PhysRevC.88.044909}{{\em Phys.
  Rev.} {\bfseries C88} no.~4, (2013) 044909},
\href{http://arxiv.org/abs/1301.4361}{{\ttfamily arXiv:1301.4361 [nucl-ex]}}.

\bibitem{Tsukada:2017llu}
K.~Tsukada {\em et~al.}, ``{First elastic electron scattering from $^{132}$Xe
  at the SCRIT facility},''
  \href{http://dx.doi.org/10.1103/PhysRevLett.118.262501}{{\em Phys. Rev.
  Lett.} {\bfseries 118} no.~26, (2017) 262501},
\href{http://arxiv.org/abs/1703.04278}{{\ttfamily arXiv:1703.04278 [nucl-ex]}}.

\bibitem{ALICE-PUBLIC-2018-003}
{\bfseries ALICE Collaboration} Collaboration, ``{Centrality determination
  using the Glauber model in Xe-Xe collisions at $\sqrt{s_{\rm NN}} = 5.44$
  TeV},''. \url{http://cds.cern.ch/record/2315401}.

\bibitem{Loizides:2017ack}
C.~Loizides, J.~Kamin, and D.~d'Enterria, ``{Precision Monte Carlo Glauber
  predictions at present and future nuclear colliders},''
\href{http://arxiv.org/abs/1710.07098}{{\ttfamily arXiv:1710.07098 [nucl-ex]}}.

\bibitem{Adam:2015kda}
{\bfseries ALICE} Collaboration, J.~Adam {\em et~al.}, ``{Centrality evolution
  of the charged-particle pseudorapidity density over a broad pseudorapidity
  range in Pb-Pb collisions at \snn\ = 2.76 TeV},''
  \href{http://dx.doi.org/10.1016/j.physletb.2015.12.082}{{\em Phys. Lett.}
  {\bfseries B754} (2016) 373--385},
\href{http://arxiv.org/abs/1509.07299}{{\ttfamily arXiv:1509.07299 [nucl-ex]}}.

\bibitem{Wang:1991hta}
X.-N. Wang and M.~Gyulassy, ``{HIJING: A Monte Carlo model for multiple jet
  production in pp, p--A and A--A collisions},''
\href{http://dx.doi.org/10.1103/PhysRevD.44.3501}{{\em Phys. Rev.} {\bfseries
  D44} (1991) 3501--3516}.

\bibitem{GEANT3}
R.~Brun, F.~Bruyant, F.~Carminati, S.~Giani, M.~Maire, A.~McPherson,
  G.~Patrick, and L.~Urban, ``{GEANT Detector Description and Simulation
  Tool},''.
\url{https://cds.cern.ch/record/1082634}.

\bibitem{ALICE-PUBLIC-2017-005}
{\bfseries ALICE} Collaboration, ``{The ALICE definition of primary
  particles},''. \url{https://cds.cern.ch/record/2270008}.

\bibitem{PhysRevLett.109.252302}
{\bfseries ALICE} Collaboration, B.~Abelev {\em et~al.}, ``{Measurement of the
  Cross Section for Electromagnetic Dissociation with Neutron Emission in Pb-Pb
  Collisions at $\sqrt{s_\mathrm{NN}}$ = 2.76 TeV},''
  \href{http://dx.doi.org/10.1103/PhysRevLett.109.252302}{{\em Phys. Rev.
  Lett.} {\bfseries 109} (2012) 252302},
\href{http://arxiv.org/abs/1203.2436}{{\ttfamily arXiv:1203.2436 [nucl-ex]}}.

\bibitem{PhysRevC.83.024913}
{\bfseries PHOBOS} Collaboration, B.~Alver {\em et~al.}, ``Charged-particle
  multiplicity and pseudorapidity distributions measured with the {PHOBOS}
  detector in {Au+Au}, {Cu+Cu}, {d+Au}, {p+p} collisions at ultrarelativistic
  energies,'' \href{http://dx.doi.org/10.1103/PhysRevC.83.024913}{{\em Phys.
  Rev.} {\bfseries C83} (2011) 024913},
  \href{http://arxiv.org/abs/1011.1940}{{\ttfamily arXiv:1011.1940 [nucl-ex]}}.

\bibitem{Abreu:2002fw}
{\bfseries NA50} Collaboration, M.~C. Abreu {\em et~al.}, ``{Scaling of charged
  particle multiplicity in {Pb-Pb} collisions at {SPS} energies},''
\href{http://dx.doi.org/10.1016/S0370-2693(02)01353-9}{{\em Phys. Lett.}
  {\bfseries B530} (2002) 43--55}.

\bibitem{Bearden:2001xw}
{\bfseries BRAHMS} Collaboration, I.~G. Bearden {\em et~al.}, ``{Charged
  particle densities from Au+Au collisions at $\sqrt{{s}_{\mathrm{NN}}}=130$
  GeV},'' \href{http://dx.doi.org/10.1016/S0370-2693(01)01333-8}{{\em Phys.
  Lett.} {\bfseries B523} (2001) 227--233},
\href{http://arxiv.org/abs/nucl-ex/0108016}{{\ttfamily arXiv:nucl-ex/0108016
  [nucl-ex]}}.

\bibitem{Bearden:2001qq}
{\bfseries BRAHMS} Collaboration, I.~G. Bearden {\em et~al.}, ``{Pseudorapidity
  distributions of charged particles from Au+Au collisions at the maximum RHIC
  energy},'' \href{http://dx.doi.org/10.1103/PhysRevLett.88.202301}{{\em Phys.
  Rev. Lett.} {\bfseries 88} (2002) 202301},
\href{http://arxiv.org/abs/nucl-ex/0112001}{{\ttfamily arXiv:nucl-ex/0112001
  [nucl-ex]}}.

\bibitem{Adcox:2000sp}
{\bfseries PHENIX} Collaboration, K.~Adcox {\em et~al.}, ``Centrality
  dependence of charged particle multiplicity in {Au-Au} collisions at \snn\ =
  130 {GeV},'' \href{http://dx.doi.org/10.1103/PhysRevLett.86.3500}{{\em Phys.
  Rev. Lett.} {\bfseries 86} (2001) 3500--3505},
\href{http://arxiv.org/abs/nucl-ex/0012008}{{\ttfamily arXiv:nucl-ex/0012008
  [nucl-ex]}}.

\bibitem{Abelev:2008ab}
{\bfseries STAR} Collaboration, B.~I. Abelev {\em et~al.}, ``Systematic
  measurements of identified particle spectra in {pp}, {d + Au} and {Au + Au}
  collisions from {STAR},''
  \href{http://dx.doi.org/10.1103/PhysRevC.79.034909}{{\em Phys. Rev.}
  {\bfseries C79} (2009) 034909},
\href{http://arxiv.org/abs/0808.2041}{{\ttfamily arXiv:0808.2041 [nucl-ex]}}.

\bibitem{Khachatryan2015143}
{\bfseries CMS} Collaboration, V.~Khachatryan {\em et~al.}, ``{Pseudorapidity
  distribution of charged hadrons in proton-proton collisions at $\sqrt{s}$ =
  13 TeV},'' \href{http://dx.doi.org/10.1016/j.physletb.2015.10.004}{{\em Phys.
  Lett.} {\bfseries B751} (2015) 143--163},
\href{http://arxiv.org/abs/1507.05915}{{\ttfamily arXiv:1507.05915 [hep-ex]}}.

\bibitem{Adam2016319}
{\bfseries ALICE} Collaboration, J.~Adam {\em et~al.}, ``{Pseudorapidity and
  transverse-momentum distributions of charged particles in protonproton
  collisions at $\sqrt s=$ 13 TeV},''
  \href{http://dx.doi.org/10.1016/j.physletb.2015.12.030}{{\em Phys. Lett.}
  {\bfseries B753} (2016) 319--329},
\href{http://arxiv.org/abs/1509.08734}{{\ttfamily arXiv:1509.08734 [nucl-ex]}}.

\bibitem{Back:2003hx}
{\bfseries PHOBOS} Collaboration, B.~B. Back {\em et~al.}, ``Pseudorapidity
  distribution of charged particles in {$d$ + Au} collisions at \snn\ = 200
  {GeV},'' \href{http://dx.doi.org/10.1103/PhysRevLett.93.082301}{{\em Phys.
  Rev. Lett.} {\bfseries 93} (2004) 082301},
\href{http://arxiv.org/abs/nucl-ex/0311009}{{\ttfamily arXiv:nucl-ex/0311009
  [nucl-ex]}}.

\bibitem{Antinori:2001qn}
{\bfseries NA57 and WA97s} Collaboration, F.~Antinori {\em et~al.},
  ``{Determination of the event centrality in the WA97 and NA57 experiments},''
\href{http://dx.doi.org/10.1088/0954-3899/27/3/317}{{\em J.Phys.} {\bfseries
  G27} (2001) 391--396}.

\bibitem{Aggarwal:2001}
{\bfseries WA98} Collaboration, M.~M. Aggarwal {\em et~al.}, ``{Scaling of
  Particle and Transverse Energy Production in $^{208}$Pb+$^{208}$Pb collisions
  at 158 A GeV},'' {\em Eur. Phys. J.} {\bfseries C18} (2001) 651--663.

\bibitem{Adare:2015bua}
{\bfseries PHENIX} Collaboration, A.~Adare {\em et~al.}, ``{Transverse energy
  production and charged-particle multiplicity at midrapidity in various
  systems from $\sqrt{s_\mathrm{NN}}=7.7$ to 200 GeV},''
  \href{http://dx.doi.org/10.1103/PhysRevC.93.024901}{{\em Phys. Rev.}
  {\bfseries C93} no.~2, (2016) 024901},
\href{http://arxiv.org/abs/1509.06727}{{\ttfamily arXiv:1509.06727 [nucl-ex]}}.

\bibitem{Adler2005}
{\bfseries PHENIX} Collaboration, S.~S. Adler {\em et~al.}, ``{Systematic
  studies of the centrality and \snn\ dependence of the
  $\mathrm{dE(T)}/\mathrm{d}\eta$ and d \dndeta\ in heavy ion collisions at
  mid-rapidity},'' \href{http://dx.doi.org/10.1103/PhysRevC.71.049901,
  10.1103/PhysRevC.71.034908}{{\em Phys. Rev.} {\bfseries C71} (2005) 034908},
  \href{http://arxiv.org/abs/nucl-ex/0409015}{{\ttfamily arXiv:nucl-ex/0409015
  [nucl-ex]}}.
[Erratum: Phys. Rev.C71,049901(2005)].

\bibitem{Werner:2007bf}
K.~Werner, ``{Core-corona separation in ultra-relativistic heavy ion
  collisions},'' \href{http://dx.doi.org/10.1103/PhysRevLett.98.152301}{{\em
  Phys. Rev. Lett.} {\bfseries 98} (2007) 152301},
\href{http://arxiv.org/abs/0704.1270}{{\ttfamily arXiv:0704.1270 [nucl-th]}}.

\bibitem{Loizides:2016djv}
C.~Loizides, ``{Glauber modeling of high-energy nuclear collisions at the
  subnucleon level},'' \href{http://dx.doi.org/10.1103/PhysRevC.94.024914}{{\em
  Phys. Rev.} {\bfseries C94} no.~2, (2016) 024914},
\href{http://arxiv.org/abs/1603.07375}{{\ttfamily arXiv:1603.07375 [nucl-ex]}}.

\bibitem{Eremin:2003qn}
S.~Eremin and S.~Voloshin, ``{Nucleon participants or quark participants?},''
  \href{http://dx.doi.org/10.1103/PhysRevC.67.064905}{{\em Phys. Rev.}
  {\bfseries C67} (2003) 064905},
\href{http://arxiv.org/abs/nucl-th/0302071}{{\ttfamily arXiv:nucl-th/0302071
  [nucl-th]}}.

\bibitem{pubNote5023}
{\bfseries ALICE} Collaboration, J.~Adam {\em et~al.}, ``{Centrality dependence
  of the charged-particle multiplicity density at midrapidity in Pb-Pb
  collisions at \snn\ = 5.02 TeV},''. \url{https://cds.cern.ch/record/2118084}.

\bibitem{Xu:2012au}
R.~Xu, W.-T. Deng, and X.-N. Wang, ``{Nuclear modification of high-pT hadron
  spectra in p+A collisions at LHC},''
  \href{http://dx.doi.org/10.1103/PhysRevC.86.051901}{{\em Phys. Rev.}
  {\bfseries C86} (2012) 051901},
\href{http://arxiv.org/abs/1204.1998}{{\ttfamily arXiv:1204.1998 [nucl-th]}}.

\bibitem{PhysRevC.83.014915}
W.-T. Deng, X.-N. Wang, and R.~Xu, ``{Hadron production in p+p, p+Pb, and Pb+Pb
  collisions with the HIJING 2.0 model at energies available at the CERN Large
  Hadron Collider},'' \href{http://dx.doi.org/10.1103/PhysRevC.83.014915}{{\em
  Phys. Rev.} {\bfseries C83} (2011) 014915},
\href{http://arxiv.org/abs/1008.1841}{{\ttfamily arXiv:1008.1841 [hep-ph]}}.

\bibitem{ASHMAN1988603}
{\bfseries European Muon} Collaboration, J.~Ashman {\em et~al.}, ``{Measurement
  of the ratios of deep inelastic muon - nucleus cross-sections on various
  nuclei compared to deuterium},''
\href{http://dx.doi.org/10.1016/0370-2693(88)91872-2}{{\em Phys. Lett.}
  {\bfseries B202} (1988) 603--610}.

\bibitem{PhysRevC.92.034906}
T.~Pierog, I.~Karpenko, J.~M. Katzy, E.~Yatsenko, and K.~Werner, ``{EPOS LHC:
  Test of collective hadronization with data measured at the CERN Large Hadron
  Collider},'' \href{http://dx.doi.org/10.1103/PhysRevC.92.034906}{{\em Phys.
  Rev.} {\bfseries C92} no.~3, (2015) 034906},
\href{http://arxiv.org/abs/1306.0121}{{\ttfamily arXiv:1306.0121 [hep-ph]}}.

\bibitem{Lin:2004en}
Z.-W. Lin, C.~M. Ko, B.-A. Li, B.~Zhang, and S.~Pal, ``{A Multi-phase transport
  model for relativistic heavy ion collisions},''
  \href{http://dx.doi.org/10.1103/PhysRevC.72.064901}{{\em Phys. Rev.}
  {\bfseries C72} (2005) 064901},
\href{http://arxiv.org/abs/nucl-th/0411110}{{\ttfamily arXiv:nucl-th/0411110
  [nucl-th]}}.

\bibitem{Xu:2011fi}
J.~Xu and C.~M. Ko, ``{Pb-Pb collisions at $\sqrt{s_\mathrm{NN}}=2.76$ TeV in a
  multiphase transport model},''
  \href{http://dx.doi.org/10.1103/PhysRevC.83.034904}{{\em Phys. Rev.}
  {\bfseries C83} (2011) 034904},
\href{http://arxiv.org/abs/1101.2231}{{\ttfamily arXiv:1101.2231 [nucl-th]}}.

\bibitem{Bierlich:2016smv}
C.~Bierlich, G.~Gustafson, and L.~L{\"o}nnblad, ``{Diffractive and
  non-diffractive wounded nucleons and final states in pA collisions},''
  \href{http://dx.doi.org/10.1007/JHEP10(2016)139}{{\em JHEP} {\bfseries 10}
  (2016) 139},
\href{http://arxiv.org/abs/1607.04434}{{\ttfamily arXiv:1607.04434 [hep-ph]}}.

\bibitem{Sjostrand:2014zea}
T.~Sj{\"o}strand, S.~Ask, J.~R. Christiansen, R.~Corke, N.~Desai, P.~Ilten,
  S.~Mrenna, S.~Prestel, C.~O. Rasmussen, and P.~Z. Skands, ``{An Introduction
  to PYTHIA 8.2},'' \href{http://dx.doi.org/10.1016/j.cpc.2015.01.024}{{\em
  Comput. Phys. Commun.} {\bfseries 191} (2015) 159--177},
\href{http://arxiv.org/abs/1410.3012}{{\ttfamily arXiv:1410.3012 [hep-ph]}}.

\bibitem{Albacete:2012xq}
J.~L. Albacete, A.~Dumitru, H.~Fujii, and Y.~Nara, ``{CGC predictions for p +
  Pb collisions at the LHC},''
  \href{http://dx.doi.org/10.1016/j.nuclphysa.2012.09.012}{{\em Nucl. Phys.}
  {\bfseries A897} (2013) 1--27},
\href{http://arxiv.org/abs/1209.2001}{{\ttfamily arXiv:1209.2001 [hep-ph]}}.

\bibitem{Albacete:2010ad}
J.~L. Albacete and A.~Dumitru, ``A model for gluon production in heavy-ion
  collisions at the {LHC} with {rcBK} unintegrated gluon densities,''
\href{http://arxiv.org/abs/1011.5161}{{\ttfamily arXiv:1011.5161 [hep-ph]}}.

\bibitem{PhysRevC.85.044920}
A.~Dumitru, D.~E. Kharzeev, E.~M. Levin, and Y.~Nara, ``{Gluon Saturation in
  $pA$ Collisions at the LHC: KLN Model Predictions For Hadron
  Multiplicities},'' \href{http://dx.doi.org/10.1103/PhysRevC.85.044920}{{\em
  Phys. Rev.} {\bfseries C85} (2012) 044920},
\href{http://arxiv.org/abs/1111.3031}{{\ttfamily arXiv:1111.3031 [hep-ph]}}.

\bibitem{Kharzeev:2001gp}
D.~Kharzeev and E.~Levin, ``{Manifestations of high density QCD in the first
  RHIC data},'' \href{http://dx.doi.org/10.1016/S0370-2693(01)01309-0}{{\em
  Phys. Lett.} {\bfseries B523} (2001) 79--87},
\href{http://arxiv.org/abs/nucl-th/0108006}{{\ttfamily arXiv:nucl-th/0108006
  [nucl-th]}}.

\bibitem{Kharzeev:2004if}
D.~Kharzeev, E.~Levin, and M.~Nardi, ``{Color glass condensate at the LHC:
  Hadron multiplicities in pp, pA and AA collisions},''
  \href{http://dx.doi.org/10.1016/j.nuclphysa.2004.10.018}{{\em Nucl. Phys.}
  {\bfseries A747} (2005) 609--629},
\href{http://arxiv.org/abs/hep-ph/0408050}{{\ttfamily arXiv:hep-ph/0408050
  [hep-ph]}}.

\bibitem{Kharzeev:2000ph}
D.~Kharzeev and M.~Nardi, ``{Hadron production in nuclear collisions at RHIC
  and high density QCD},''
  \href{http://dx.doi.org/10.1016/S0370-2693(01)00457-9}{{\em Phys. Lett.}
  {\bfseries B507} (2001) 121--128},
\href{http://arxiv.org/abs/nucl-th/0012025}{{\ttfamily arXiv:nucl-th/0012025
  [nucl-th]}}.

\bibitem{Armesto:2004ud}
N.~Armesto, C.~A. Salgado, and U.~A. Wiedemann, ``{Relating high-energy
  lepton-hadron, proton-nucleus and nucleus-nucleus collisions through
  geometric scaling},''
  \href{http://dx.doi.org/10.1103/PhysRevLett.94.022002}{{\em Phys. Rev. Lett.}
  {\bfseries 94} (2005) 022002},
\href{http://arxiv.org/abs/hep-ph/0407018}{{\ttfamily arXiv:hep-ph/0407018
  [hep-ph]}}.

\bibitem{Schenke:2013dpa}
B.~Schenke, P.~Tribedy, and R.~Venugopalan, ``{Multiplicity distributions in
  p+p, p+A and A+A collisions from Yang-Mills dynamics},''
  \href{http://dx.doi.org/10.1103/PhysRevC.89.024901}{{\em Phys. Rev.}
  {\bfseries C89} no.~2, (2014) 024901},
\href{http://arxiv.org/abs/1311.3636}{{\ttfamily arXiv:1311.3636 [hep-ph]}}.

\bibitem{Schenke:2012wb}
B.~Schenke, P.~Tribedy, and R.~Venugopalan, ``{Fluctuating Glasma initial
  conditions and flow in heavy ion collisions},''
  \href{http://dx.doi.org/10.1103/PhysRevLett.108.252301}{{\em Phys. Rev.
  Lett.} {\bfseries 108} (2012) 252301},
\href{http://arxiv.org/abs/1202.6646}{{\ttfamily arXiv:1202.6646 [nucl-th]}}.

\bibitem{Niemi:2015qia}
H.~Niemi, K.~J. Eskola, and R.~Paatelainen, ``{Event-by-event fluctuations in a
  perturbative QCD + saturation + hydrodynamics model: Determining QCD matter
  shear viscosity in ultrarelativistic heavy-ion collisions},''
  \href{http://dx.doi.org/10.1103/PhysRevC.93.024907}{{\em Phys. Rev.}
  {\bfseries C93} no.~2, (2016) 024907},
\href{http://arxiv.org/abs/1505.02677}{{\ttfamily arXiv:1505.02677 [hep-ph]}}.

\bibitem{Niemi:2015voa}
H.~Niemi, K.~J. Eskola, R.~Paatelainen, and K.~Tuominen, ``{Predictions for
  5.023 TeV Pb + Pb collisions at the CERN Large Hadron Collider},''
  \href{http://dx.doi.org/10.1103/PhysRevC.93.014912}{{\em Phys. Rev.}
  {\bfseries C93} no.~1, (2016) 014912},
\href{http://arxiv.org/abs/1511.04296}{{\ttfamily arXiv:1511.04296 [hep-ph]}}.

\bibitem{Eskola:2017bup}
K.~J. Eskola, H.~Niemi, R.~Paatelainen, and K.~Tuominen, ``{Predictions for
  multiplicities and flow harmonics in 5.44 TeV Xe+Xe collisions at the CERN
  Large Hadron Collider},''
  \href{http://dx.doi.org/10.1103/PhysRevC.97.034911}{{\em Phys. Rev.}
  {\bfseries C97} no.~3, (2018) 034911},
\href{http://arxiv.org/abs/1711.09803}{{\ttfamily arXiv:1711.09803 [hep-ph]}}.

\bibitem{Bass:2017zyn}
S.~A. Bass, J.~E. Bernhard, and J.~S. Moreland, ``{Determination of
  Quark-Gluon-Plasma Parameters from a Global Bayesian Analysis},''
  \href{http://dx.doi.org/10.1016/j.nuclphysa.2017.05.052}{{\em Nucl. Phys.}
  {\bfseries A967} (2017) 67--73},
\href{http://arxiv.org/abs/1704.07671}{{\ttfamily arXiv:1704.07671 [nucl-th]}}.

\bibitem{Moreland:2014oya}
J.~S. Moreland, J.~E. Bernhard, and S.~A. Bass, ``{Alternative ansatz to
  wounded nucleon and binary collision scaling in high-energy nuclear
  collisions},'' \href{http://dx.doi.org/10.1103/PhysRevC.92.011901}{{\em Phys.
  Rev.} {\bfseries C92} no.~1, (2015) 011901},
\href{http://arxiv.org/abs/1412.4708}{{\ttfamily arXiv:1412.4708 [nucl-th]}}.

\bibitem{Shen:2014vra}
C.~Shen, Z.~Qiu, H.~Song, J.~Bernhard, S.~Bass, and U.~Heinz, ``{The
  iEBE-VISHNU code package for relativistic heavy-ion collisions},''
  \href{http://dx.doi.org/10.1016/j.cpc.2015.08.039}{{\em Comput. Phys.
  Commun.} {\bfseries 199} (2016) 61--85},
\href{http://arxiv.org/abs/1409.8164}{{\ttfamily arXiv:1409.8164 [nucl-th]}}.

\bibitem{PhysRevC.70.021902}
{\bfseries PHOBOS} Collaboration, B.~B. Back {\em et~al.}, ``{Collision
  geometry scaling of Au+Au pseudorapidity density from s(NN)**(1/2) = 19.6-GeV
  to 200-GeV},'' \href{http://dx.doi.org/10.1103/PhysRevC.70.021902}{{\em Phys.
  Rev.} {\bfseries C70} (2004) 021902},
\href{http://arxiv.org/abs/nucl-ex/0405027}{{\ttfamily arXiv:nucl-ex/0405027
  [nucl-ex]}}.

\bibitem{Aamodt:2010pb}
{\bfseries ALICE} Collaboration, K.~Aamodt {\em et~al.}, ``{Charged--particle
  multiplicity density at mid--rapidity in central \PbPb\ collisions at \snn\ =
  2.76 TeV},'' \href{http://dx.doi.org/10.1103/PhysRevLett.105.252301}{{\em
  Phys. Rev. Lett.} {\bfseries 105} (2010) 252301},
\href{http://arxiv.org/abs/1011.3916}{{\ttfamily arXiv:1011.3916 [nucl-ex]}}.

\bibitem{Rezaeian:2012ji}
A.~H. Rezaeian, M.~Siddikov, M.~Van~de Klundert, and R.~Venugopalan,
  ``{Analysis of combined HERA data in the Impact-Parameter dependent
  Saturation model},'' \href{http://dx.doi.org/10.1103/PhysRevD.87.034002}{{\em
  Phys. Rev.} {\bfseries D87} no.~3, (2013) 034002},
\href{http://arxiv.org/abs/1212.2974}{{\ttfamily arXiv:1212.2974 [hep-ph]}}.

\bibitem{PhysRevLett.102.142301}
{\bfseries PHOBOS} Collaboration, B.~Alver {\em et~al.}, ``{System size, energy
  and centrality dependence of pseudorapidity distributions of charged
  particles in relativistic heavy ion collisions},''
  \href{http://dx.doi.org/10.1103/PhysRevLett.102.142301}{{\em Phys. Rev.
  Lett.} {\bfseries 102} (2009) 142301},
\href{http://arxiv.org/abs/0709.4008}{{\ttfamily arXiv:0709.4008 [nucl-ex]}}.

\end{thebibliography}\endgroup

\newpage
\appendix
\section{The ALICE Collaboration}
\label{app:collab}

\begingroup
\small
\begin{flushleft}
S.~Acharya\Irefn{org138}\And 
F.T.-.~Acosta\Irefn{org22}\And 
D.~Adamov\'{a}\Irefn{org94}\And 
J.~Adolfsson\Irefn{org81}\And 
M.M.~Aggarwal\Irefn{org98}\And 
G.~Aglieri Rinella\Irefn{org36}\And 
M.~Agnello\Irefn{org33}\And 
N.~Agrawal\Irefn{org49}\And 
Z.~Ahammed\Irefn{org138}\And 
S.U.~Ahn\Irefn{org77}\And 
S.~Aiola\Irefn{org143}
% ...
\And
\renewcommand\labelenumi{\textsuperscript{\theenumi}~}

\section*{Affiliation notes}
\renewcommand\theenumi{\roman{enumi}}
\begin{Authlist}
\item \Adef{org*}Deceased
\item \Adef{orgI}Dipartimento DET del Politecnico di Torino, Turin, Italy
\item \Adef{orgII}M.V. Lomonosov Moscow State University, D.V. Skobeltsyn Institute of Nuclear, Physics, Moscow, Russia
\item \Adef{orgIII}Department of Applied Physics, Aligarh Muslim University, Aligarh, India
\item \Adef{orgIV}Institute of Theoretical Physics, University of Wroclaw, Poland
\end{Authlist}

\section*{Collaboration Institutes}
\renewcommand\theenumi{\arabic{enumi}~}
\begin{Authlist}
\item \Idef{org1}A.I. Alikhanyan National Science Laboratory (Yerevan Physics Institute) Foundation, Yerevan, Armenia
\item \Idef{org2}Benem\'{e}rita Universidad Aut\'{o}noma de Puebla, Puebla, Mexico
\item \Idef{org3}Bogolyubov Institute for Theoretical Physics, National Academy of Sciences of Ukraine, Kiev, Ukraine
\item \Idef{org4}Bose Institute, Department of Physics  and Centre for Astroparticle Physics and Space Science (CAPSS), Kolkata, India
\item \Idef{org5}Budker Institute for Nuclear Physics, Novosibirsk, Russia
\item \Idef{org6}California Polytechnic State University, San Luis Obispo, California, United States
\item \Idef{org7}Central China Normal University, Wuhan, China
\item \Idef{org8}Centre de Calcul de l'IN2P3, Villeurbanne, Lyon, France
\item \Idef{org9}Centro de Aplicaciones Tecnol\'{o}gicas y Desarrollo Nuclear (CEADEN), Havana, Cuba
\item \Idef{org10}Centro de Investigaci\'{o}n y de Estudios Avanzados (CINVESTAV), Mexico City and M\'{e}rida, Mexico
\item \Idef{org11}Centro Fermi - Museo Storico della Fisica e Centro Studi e Ricerche ``Enrico Fermi', Rome, Italy
\item \Idef{org12}Chicago State University, Chicago, Illinois, United States
\item \Idef{org13}China Institute of Atomic Energy, Beijing, China
\item \Idef{org14}Chonbuk National University, Jeonju, Republic of Korea
\item \Idef{org15}Comenius University Bratislava, Faculty of Mathematics, Physics and Informatics, Bratislava, Slovakia
\item \Idef{org16}COMSATS Institute of Information Technology (CIIT), Islamabad, Pakistan
\item \Idef{org17}Creighton University, Omaha, Nebraska, United States
\item \Idef{org18}Department of Physics, Aligarh Muslim University, Aligarh, India
\item \Idef{org19}Department of Physics, Ohio State University, Columbus, Ohio, United States
\item \Idef{org20}Department of Physics, Pusan National University, Pusan, Republic of Korea
\item \Idef{org21}Department of Physics, Sejong University, Seoul, Republic of Korea
\item \Idef{org22}Department of Physics, University of California, Berkeley, California, United States
\item \Idef{org23}Department of Physics, University of Oslo, Oslo, Norway
\item \Idef{org24}Department of Physics and Technology, University of Bergen, Bergen, Norway
\item \Idef{org25}Dipartimento di Fisica dell'Universit\`{a} 'La Sapienza' and Sezione INFN, Rome, Italy
\item \Idef{org26}Dipartimento di Fisica dell'Universit\`{a} and Sezione INFN, Cagliari, Italy
\item \Idef{org27}Dipartimento di Fisica dell'Universit\`{a} and Sezione INFN, Trieste, Italy
\item \Idef{org28}Dipartimento di Fisica dell'Universit\`{a} and Sezione INFN, Turin, Italy
\item \Idef{org29}Dipartimento di Fisica e Astronomia dell'Universit\`{a} and Sezione INFN, Bologna, Italy
\item \Idef{org30}Dipartimento di Fisica e Astronomia dell'Universit\`{a} and Sezione INFN, Catania, Italy
\item \Idef{org31}Dipartimento di Fisica e Astronomia dell'Universit\`{a} and Sezione INFN, Padova, Italy
\item \Idef{org32}Dipartimento di Fisica `E.R.~Caianiello' dell'Universit\`{a} and Gruppo Collegato INFN, Salerno, Italy
\item \Idef{org33}Dipartimento DISAT del Politecnico and Sezione INFN, Turin, Italy
\item \Idef{org34}Dipartimento di Scienze e Innovazione Tecnologica dell'Universit\`{a} del Piemonte Orientale and INFN Sezione di Torino, Alessandria, Italy
\item \Idef{org35}Dipartimento Interateneo di Fisica `M.~Merlin' and Sezione INFN, Bari, Italy
\item \Idef{org36}European Organization for Nuclear Research (CERN), Geneva, Switzerland
\item \Idef{org37}Faculty of Electrical Engineering, Mechanical Engineering and Naval Architecture, University of Split, Split, Croatia
\item \Idef{org38}Faculty of Engineering and Science, Western Norway University of Applied Sciences, Bergen, Norway
\item \Idef{org39}Faculty of Nuclear Sciences and Physical Engineering, Czech Technical University in Prague, Prague, Czech Republic
\item \Idef{org40}Faculty of Science, P.J.~\v{S}af\'{a}rik University, Ko\v{s}ice, Slovakia
\item \Idef{org41}Frankfurt Institute for Advanced Studies, Johann Wolfgang Goethe-Universit\"{a}t Frankfurt, Frankfurt, Germany
\item \Idef{org42}Gangneung-Wonju National University, Gangneung, Republic of Korea
\item \Idef{org43}Gauhati University, Department of Physics, Guwahati, India
\item \Idef{org44}Helmholtz-Institut f\"{u}r Strahlen- und Kernphysik, Rheinische Friedrich-Wilhelms-Universit\"{a}t Bonn, Bonn, Germany
\item \Idef{org45}Helsinki Institute of Physics (HIP), Helsinki, Finland
\item \Idef{org46}Hiroshima University, Hiroshima, Japan
\item \Idef{org47}Hochschule Worms, Zentrum  f\"{u}r Technologietransfer und Telekommunikation (ZTT), Worms, Germany
\item \Idef{org48}Horia Hulubei National Institute of Physics and Nuclear Engineering, Bucharest, Romania
\item \Idef{org49}Indian Institute of Technology Bombay (IIT), Mumbai, India
\item \Idef{org50}Indian Institute of Technology Indore, Indore, India
\item \Idef{org51}Indonesian Institute of Sciences, Jakarta, Indonesia
\item \Idef{org52}INFN, Laboratori Nazionali di Frascati, Frascati, Italy
\item \Idef{org53}INFN, Sezione di Bari, Bari, Italy
\item \Idef{org54}INFN, Sezione di Bologna, Bologna, Italy
\item \Idef{org55}INFN, Sezione di Cagliari, Cagliari, Italy
\item \Idef{org56}INFN, Sezione di Catania, Catania, Italy
\item \Idef{org57}INFN, Sezione di Padova, Padova, Italy
\item \Idef{org58}INFN, Sezione di Roma, Rome, Italy
\item \Idef{org59}INFN, Sezione di Torino, Turin, Italy
\item \Idef{org60}INFN, Sezione di Trieste, Trieste, Italy
\item \Idef{org61}Inha University, Incheon, Republic of Korea
\item \Idef{org62}Institut de Physique Nucl\'{e}aire d'Orsay (IPNO), Institut National de Physique Nucl\'{e}aire et de Physique des Particules (IN2P3/CNRS), Universit\'{e} de Paris-Sud, Universit\'{e} Paris-Saclay, Orsay, France
\item \Idef{org63}Institute for Nuclear Research, Academy of Sciences, Moscow, Russia
\item \Idef{org64}Institute for Subatomic Physics, Utrecht University/Nikhef, Utrecht, Netherlands
\item \Idef{org65}Institute for Theoretical and Experimental Physics, Moscow, Russia
\item \Idef{org66}Institute of Experimental Physics, Slovak Academy of Sciences, Ko\v{s}ice, Slovakia
\item \Idef{org67}Institute of Physics, Bhubaneswar, India
\item \Idef{org68}Institute of Physics of the Czech Academy of Sciences, Prague, Czech Republic
\item \Idef{org69}Institute of Space Science (ISS), Bucharest, Romania
\item \Idef{org70}Institut f\"{u}r Kernphysik, Johann Wolfgang Goethe-Universit\"{a}t Frankfurt, Frankfurt, Germany
\item \Idef{org71}Instituto de Ciencias Nucleares, Universidad Nacional Aut\'{o}noma de M\'{e}xico, Mexico City, Mexico
\item \Idef{org72}Instituto de F\'{i}sica, Universidade Federal do Rio Grande do Sul (UFRGS), Porto Alegre, Brazil
\item \Idef{org73}Instituto de F\'{\i}sica, Universidad Nacional Aut\'{o}noma de M\'{e}xico, Mexico City, Mexico
\item \Idef{org74}iThemba LABS, National Research Foundation, Somerset West, South Africa
\item \Idef{org75}Johann-Wolfgang-Goethe Universit\"{a}t Frankfurt Institut f\"{u}r Informatik, Fachbereich Informatik und Mathematik, Frankfurt, Germany
\item \Idef{org76}Joint Institute for Nuclear Research (JINR), Dubna, Russia
\item \Idef{org77}Korea Institute of Science and Technology Information, Daejeon, Republic of Korea
\item \Idef{org78}KTO Karatay University, Konya, Turkey
\item \Idef{org79}Laboratoire de Physique Subatomique et de Cosmologie, Universit\'{e} Grenoble-Alpes, CNRS-IN2P3, Grenoble, France
\item \Idef{org80}Lawrence Berkeley National Laboratory, Berkeley, California, United States
\item \Idef{org81}Lund University Department of Physics, Division of Particle Physics, Lund, Sweden
\item \Idef{org82}Nagasaki Institute of Applied Science, Nagasaki, Japan
\item \Idef{org83}Nara Women{'}s University (NWU), Nara, Japan
\item \Idef{org84}National and Kapodistrian University of Athens, School of Science, Department of Physics , Athens, Greece
\item \Idef{org85}National Centre for Nuclear Research, Warsaw, Poland
\item \Idef{org86}National Institute of Science Education and Research, HBNI, Jatni, India
\item \Idef{org87}National Nuclear Research Center, Baku, Azerbaijan
\item \Idef{org88}National Research Centre Kurchatov Institute, Moscow, Russia
\item \Idef{org89}Niels Bohr Institute, University of Copenhagen, Copenhagen, Denmark
\item \Idef{org90}Nikhef, National institute for subatomic physics, Amsterdam, Netherlands
\item \Idef{org91}NRC Kurchatov Institute IHEP, Protvino, Russia
\item \Idef{org92}NRNU Moscow Engineering Physics Institute, Moscow, Russia
\item \Idef{org93}Nuclear Physics Group, STFC Daresbury Laboratory, Daresbury, United Kingdom
\item \Idef{org94}Nuclear Physics Institute of the Czech Academy of Sciences, \v{R}e\v{z} u Prahy, Czech Republic
\item \Idef{org95}Oak Ridge National Laboratory, Oak Ridge, Tennessee, United States
\item \Idef{org96}Petersburg Nuclear Physics Institute, Gatchina, Russia
\item \Idef{org97}Physics department, Faculty of science, University of Zagreb, Zagreb, Croatia
\item \Idef{org98}Physics Department, Panjab University, Chandigarh, India
\item \Idef{org99}Physics Department, University of Jammu, Jammu, India
\item \Idef{org100}Physics Department, University of Rajasthan, Jaipur, India
\item \Idef{org101}Physikalisches Institut, Eberhard-Karls-Universit\"{a}t T\"{u}bingen, T\"{u}bingen, Germany
\item \Idef{org102}Physikalisches Institut, Ruprecht-Karls-Universit\"{a}t Heidelberg, Heidelberg, Germany
\item \Idef{org103}Physik Department, Technische Universit\"{a}t M\"{u}nchen, Munich, Germany
\item \Idef{org104}Research Division and ExtreMe Matter Institute EMMI, GSI Helmholtzzentrum f\"ur Schwerionenforschung GmbH, Darmstadt, Germany
\item \Idef{org105}Rudjer Bo\v{s}kovi\'{c} Institute, Zagreb, Croatia
\item \Idef{org106}Russian Federal Nuclear Center (VNIIEF), Sarov, Russia
\item \Idef{org107}Saha Institute of Nuclear Physics, Kolkata, India
\item \Idef{org108}School of Physics and Astronomy, University of Birmingham, Birmingham, United Kingdom
\item \Idef{org109}Secci\'{o}n F\'{\i}sica, Departamento de Ciencias, Pontificia Universidad Cat\'{o}lica del Per\'{u}, Lima, Peru
\item \Idef{org110}Shanghai Institute of Applied Physics, Shanghai, China
\item \Idef{org111}Stefan Meyer Institut f\"{u}r Subatomare Physik (SMI), Vienna, Austria
\item \Idef{org112}SUBATECH, IMT Atlantique, Universit\'{e} de Nantes, CNRS-IN2P3, Nantes, France
\item \Idef{org113}Suranaree University of Technology, Nakhon Ratchasima, Thailand
\item \Idef{org114}Technical University of Ko\v{s}ice, Ko\v{s}ice, Slovakia
\item \Idef{org115}Technische Universit\"{a}t M\"{u}nchen, Excellence Cluster 'Universe', Munich, Germany
\item \Idef{org116}The Henryk Niewodniczanski Institute of Nuclear Physics, Polish Academy of Sciences, Cracow, Poland
\item \Idef{org117}The University of Texas at Austin, Austin, Texas, United States
\item \Idef{org118}Universidad Aut\'{o}noma de Sinaloa, Culiac\'{a}n, Mexico
\item \Idef{org119}Universidade de S\~{a}o Paulo (USP), S\~{a}o Paulo, Brazil
\item \Idef{org120}Universidade Estadual de Campinas (UNICAMP), Campinas, Brazil
\item \Idef{org121}Universidade Federal do ABC, Santo Andre, Brazil
\item \Idef{org122}University College of Southeast Norway, Tonsberg, Norway
\item \Idef{org123}University of Cape Town, Cape Town, South Africa
\item \Idef{org124}University of Houston, Houston, Texas, United States
\item \Idef{org125}University of Jyv\"{a}skyl\"{a}, Jyv\"{a}skyl\"{a}, Finland
\item \Idef{org126}University of Liverpool, Department of Physics Oliver Lodge Laboratory , Liverpool, United Kingdom
\item \Idef{org127}University of Tennessee, Knoxville, Tennessee, United States
\item \Idef{org128}University of the Witwatersrand, Johannesburg, South Africa
\item \Idef{org129}University of Tokyo, Tokyo, Japan
\item \Idef{org130}University of Tsukuba, Tsukuba, Japan
\item \Idef{org131}Universit\'{e} Clermont Auvergne, CNRS/IN2P3, LPC, Clermont-Ferrand, France
\item \Idef{org132}Universit\'{e} de Lyon, Universit\'{e} Lyon 1, CNRS/IN2P3, IPN-Lyon, Villeurbanne, Lyon, France
\item \Idef{org133}Universit\'{e} de Strasbourg, CNRS, IPHC UMR 7178, F-67000 Strasbourg, France, Strasbourg, France
\item \Idef{org134} Universit\'{e} Paris-Saclay Centre d¿\'Etudes de Saclay (CEA), IRFU, Department de Physique Nucl\'{e}aire (DPhN), Saclay, France
\item \Idef{org135}Universit\`{a} degli Studi di Pavia, Pavia, Italy
\item \Idef{org136}Universit\`{a} di Brescia, Brescia, Italy
\item \Idef{org137}V.~Fock Institute for Physics, St. Petersburg State University, St. Petersburg, Russia
\item \Idef{org138}Variable Energy Cyclotron Centre, Kolkata, India
\item \Idef{org139}Warsaw University of Technology, Warsaw, Poland
\item \Idef{org140}Wayne State University, Detroit, Michigan, United States
\item \Idef{org141}Westf\"{a}lische Wilhelms-Universit\"{a}t M\"{u}nster, Institut f\"{u}r Kernphysik, M\"{u}nster, Germany
\item \Idef{org142}Wigner Research Centre for Physics, Hungarian Academy of Sciences, Budapest, Hungary
\item \Idef{org143}Yale University, New Haven, Connecticut, United States
\item \Idef{org144}Yonsei University, Seoul, Republic of Korea
\end{Authlist}
\endgroup
\end{document}